\documentclass[prd,notitleapage,aps,onecolumn,10pt]{revtex4-2} 
\usepackage{natbib}
\usepackage{hyperref}
\hypersetup{colorlinks=true, citecolor=blue, urlcolor=blue, linkcolor=violet}
\usepackage{graphicx}
\usepackage{physics}
\usepackage{array}
\usepackage{amsmath,amssymb}
\usepackage{mathrsfs}
\usepackage{dsfont}
\usepackage[T1]{fontenc}
\usepackage{inputenc}
\usepackage{babel}
\usepackage{multirow}
\usepackage{longtable, array, booktabs}
\usepackage[table]{xcolor}
\usepackage{float}
\usepackage{comment}

\allowdisplaybreaks

\begin{document}
\title{Hamiltonian Flow Equations in Daubechies Wavelet Basis}
\author{Mrinmoy Basak}
\email[Electronic Address: ]{mrinmoy.263009@gmail.com}
\author{Raghunath Ratabole}
\email[Electronic Address: ]{ratabole@goa.bits-pilani.ac.in}
\affiliation{Department of Physics, BITS-Pilani KK Birla Goa Campus,
Zuarinagar 403726, Goa, India}
\begin{abstract}
We study the low energy dynamics of a system of two coupled real scalar fields in 1+1 dimensions using the flow equation approach of Similarity Renormalization Group (SRG) in a wavelet basis. This paper presents an extension of the work by Michlin and Polyzou \cite{PhysRevD.95.094501} at one resolution higher. We also present the analysis of a model of two scalar fields coupled through a generally quadratic interaction in $1+1$ dimensions using wavelet-based flow equations. We demonstrate that the specifically chosen generator flows the Hamiltonian into a block diagonal form with each diagonal block being associated with a fixed resolution. The wavelet basis is known to transform the scalar field theory into a model of coupled localized oscillators, each of which is labelled by location and resolution indices. The chosen interaction represents the coupling between two types of oscillators at the same location and resolution index. There is no coupling between oscillators across locations and resolutions. We show that wavelet-based flow equations carry out scale separation while maintaining the interactions between the two scalar fields at each resolution. The effective Hamiltonian associated with the coarsest resolution is shown to correctly reproduce the normal mode frequencies of the model.
\end{abstract}
\maketitle
\section{Introduction}

The use of wavelets for nonperturbative analysis of quantum field theories (QFTs) first emerged in a 1965 work by Wilson titled "Model Hamiltonians for Local Quantum Field Theory" \cite{PhysRev.140.B445}. In this pioneering work, Wilson explained how a divergent bare coupling constant may emerge in a continuum relativistic quantum field theory with finite physical predictions using what is known today as the Wilsonian approach to renormalization. This approach was developed using the theory of a relativistic complex scalar field (charged pion) interacting with an external quantum two-level source (the nucleon). One of his aims was to solve the Hamiltonian eigenvalue problem of relativistic QFTs using quantum mechanical techniques that go beyond perturbation theory. In order to establish the validity of these methods, it was important that one could make qualitative order of magnitude estimates of various quantities in the process of computation. This was done using phase space analysis, in which the quantum field was resolved into localized oscillator variables using a complete set of "wavepacket" basis functions. These wavepacket functions were chosen to have specific localization characteristics, and in the absence of an explicit construction, their existence was assumed. The wavepacket functions had properties akin to those of the scaling basis functions of the wavelet theory, which was to take its modern form a couple of decades later. The importance of wavelets for nonperturbative analysis of quantum chromodynamics (QCD) was re-emphasized in an approach pioneered by Wilson et al \cite{PhysRevD.49.6720}. This work outlined a proposal for the computation of bound states in QCD by combining together three key elements: the light front formulation of field theory, a wavelet representation of the fields and the Similarity Renormalization Group (SRG). The formulation of SRG by Wilson and Glazek \cite{PhysRevD.48.5863,PhysRevD.49.4214} and Wegner \cite{https://doi.org/10.1002/andp.19945060203} was a precursor to this proposal. The wavelet analysis presented in this proposal \cite{PhysRev.140.B445}, while qualitative in nature, depended only on the localized characteristics of wavelets and not on their specific type and form. The SRG approach matured into the Renormalization Group Procedure for Effective Particles (RGPEP) but did so without using wavelets. The RGPEP approach \cite{PhysRevD.85.125018} is a specific implementation of SRG, often used for the nonperturbative analysis of light front field theories formulated in the Fourier domain. Michlin and Polyzou were the first to study the flow equation implementation of SRG in the context of a $1+1$ dimensional real scalar field theory quantized in instant form \cite{PhysRevD.95.094501}. They showed the scale decomposition of the degrees of freedom using the flow equation approach. Polyzou formulated the wavelet representation of light front field theory in the context of a real scalar field \cite{PhysRevD.101.096004}.  In order to push forward the agenda of solving QCD as envisaged by Wilson in \cite{PhysRev.140.B445}, more studies on the implementation of SRG on wavelet-based field theories are required. In this work, we present an extension of the study \cite{PhysRevD.95.094501} for a quantum field theory with quadratic interaction.
 
In the standard approach to quantizing relativistic field theories, the quantum field is decomposed into its Fourier modes.  The Fourier basis functions are spread out in direct space while having extreme localisation in reciprocal space. The kinetic energy and the mass term in the Hamiltonian can be written as an uncoupled sum over the Fourier modes. On the other hand, the interaction terms in the Hamiltonian have both self-mode and inter-mode coupling terms consistent with momentum (or wave vector conservation). While this approach works well perturbatively, in the strong coupling regime, one would like to limit the extent of mode coupling in the interaction terms. This can be done by working with "wavepacket" basis functions that have approximately finite support in both direct and reciprocal space. The study of such functions constitutes the mathematical subject of wavelet analysis. The term "wavelet" is a translation of the French word "ondelette," meaning "small wave", that was first introduced by Jean Morlet and Alex Grossman \cite{doi:10.1190/1.1441328}. Their work laid the foundation for wavelet theory, which has become a crucial tool in signal processing \cite{192463,adelson1987orthogonal,rioul1991wavelets,chui1997wavelets,kovacevic2013fourier,unser1992polynomial,stephane1999wavelet,abbate1997signal,cotronei1998multiwavelet,kingsbury1998wavelet,sundararajan2016discrete} and has found applications across various disciplines of science \cite{gao2010wavelets,aldroubi1996wavelets,streit201304,10.1190/1.1441328,DREMIN2002599,PhysRevLett.80.1544,PhysRevLett.81.2388}. 

Two approaches to wavelet-based QFT, seen in the literature, decompose the quantum field into wavepacket-like modes using either a continuous wavelet transform or a discrete wavelet transform. Both approaches, discrete and continuous, have investigated regularization, renormalization, and gauge invariance in QFTs \cite{kessler2003waveletnotes,Polyzou_2018,MichlinTracieL2017Uwbt,Federbush_1995,BEST2000848,PhysRevLett.116.140403,Tomboulis2021,PhysRevD.106.036025,thiemann2022renormalisationwaveletsdirichletshannonkernels,best1994variationaldescriptionstatisticalfield,HALLIDAY1995414,10.1063/1.1543582,Polyzou_2024,Polyzou2014,PhysRevD.87.116011,PhysRevD.95.094501,Neuberger2018,Altaisky_2007,albeverio2010remarkgaugeinvariancewaveletbased,Altaisky_2018,PhysRevD.88.025015,Altaiskii2013,Altaisky:2015zca,PhysRevD.101.096004,polyzou2020lightfrontquantummechanicsquantum}.  In the discrete wavelet approach, using Daubechies wavelet basis of order $K$, the quantum field is spatially decomposed into modes labelled by discrete indices. Daubechies wavelets, first introduced by Ingrid Daubechies \cite{https://doi.org/10.1002/cpa.3160410705,doi:10.1137/1.9781611970104}, are a family of orthonormal and compactly supported wavelets. The discrete mode indices characterize the location and the length scale (or resolution) underlying the mode. Expressed in terms of the wavelet modes, the QFT Hamiltonian will have a limited amount of inter-mode coupling in both kinetic and interaction terms, as originally envisioned by Wilson. The extent of this inter-mode coupling will be determined by the order $K$ of Daubechies wavelets. An experimental setup to probe and observe the physical system under study occupies a specific volume in space with a limit on the energy of the probe or resolution of the observation.  The continuum quantum field theory that is used to model the physical system can be truncated by volume and resolution as is natural to the experimental context. Set within the framework of multiresolution analysis (MRA) \cite{LI201414}, the wavelet based methods provide a systematic framework to improve the truncation in volume and resolution.

The use of Daubechies wavelets for analysis of problems in QFT was advocated by Bulut and Polyzou \cite{PhysRevD.87.116011} by implementing wavelet based canonical quantization while studying various aspects of Poincare invariance, renormalization group and gauge invariance from a wavelet perspective. There is earlier work using Daubechies wavelets in the context of statistical field theories \cite{best1994variationaldescriptionstatisticalfield} and classical spin models \cite{BEST2000848}. The discrete wavelet approach has been leveraged to propose a quantum simulation of a self-interacting real scalar field theory \cite{PhysRevD.106.036025} and in establishing an exact holographic mapping in free fermionic and bosonic systems \cite{qi2013exactholographicmappingemergent,PhysRevA.92.032315}. Flow equations methods were tested on $(1+1)$ dimensional scalar field theory as a means to decouple short-distance degrees of freedom \cite{PhysRevD.95.094501,MichlinTracieL2017Uwbt,Polyzou2018}. The discrete canonical structure of the truncated theory provides an avenue for the construction of the real time path integral formulation of QFT \cite{Polyzou_2024}. The construction of a continuum massive free field theory as the scaling limit of Hamiltonian lattice systems using a discrete wavelet approach was reported in \cite{PhysRevLett.127.230601}. These multi-scale methods have also been used to establish \cite{PhysRevD.106.036025} the well-known results of scale-dependent entanglement entropy and renormalization of correlations in the ground state of a $(1+1)$ dimensional bosonic and fermionic free field theory. A wavelet based multiscale entanglement renormalization ansatz (wMERA) for continuum free scalar field theory in one spatial dimension was proposed by Alves \cite{Alves2024}.

In this paper, we extend the work of Polyzou and Michlin \cite{PhysRevD.95.094501} in two ways. Firstly, the flow equations are applied to Daubechies wavelet-based free scalar field theory in $1+1$ dimensions at one resolution higher than previously reported. The resulting decoupling of multiple resolution sectors is demonstrated. Secondly, we study the theory of two real scalar fields with a generally quadratic interaction. We show that the flow equations filter the interaction terms by resolution. The normal mode frequencies predicted by the effective Hamiltonian of the lowest resolution sector are compared to those of the exact truncated Hamiltonian to demonstrate the inclusion of high-resolution effects into the Hamiltonian of the low-resolution sector.



The structure of this paper is as follows: In Sec. \ref{sec:Introduction_to_the_Daubechies_Wavelet_basis}, we introduce Daubechies wavelets and review their properties, which will be used throughout the rest of this paper. After describing the flow equation method in Sec. \ref{sec:Flow-equation to separate scales and the coupling between two fields}, we analyse the $1+1$ dimensional wavelet-based free scalar field theory using the flow-equation method. Sec. \ref{sec:model_of_two_interacting_scalar_fields} and Sec. \ref{sec:Canonical_Quantization_in_wavelet_basis} presents the quantization of two real scalar fields with a generally quadratic interaction in Fourier and wavelet basis respectively. In Sec. \ref{sec:analysis}, we analyse this model from a flow equation perspective and demonstrate the resolution decomposition of the interaction terms. The concluding remarks are presented in Sec. \ref{sec:conclusion}.

\section{Introduction to the Daubechies Wavelet basis}
\label{sec:Introduction_to_the_Daubechies_Wavelet_basis}
In this section, we briefly review the construction of the Daubechies wavelet basis and its properties, which will be used in the rest of the paper. We refer the reader to \cite{kessler2003waveletnotes,PhysRevD.107.036015,PhysRevD.87.116011,PhysRevD.95.094501,doi:10.1137/1.9781611970104,388960,https://doi.org/10.1002/cpa.3160410705} for a comprehensive introduction to the subject of discrete wavelets. 

\begin{figure}[hbt]
    \centering
    \caption{Any generic function $f(x)$ (blue color). The action of (a) $\hat{D}$ and (b) $\hat{T}$ operator on that function (red color)}
    \label{fig:D_and_T_operations}
    \includegraphics[scale=0.35]{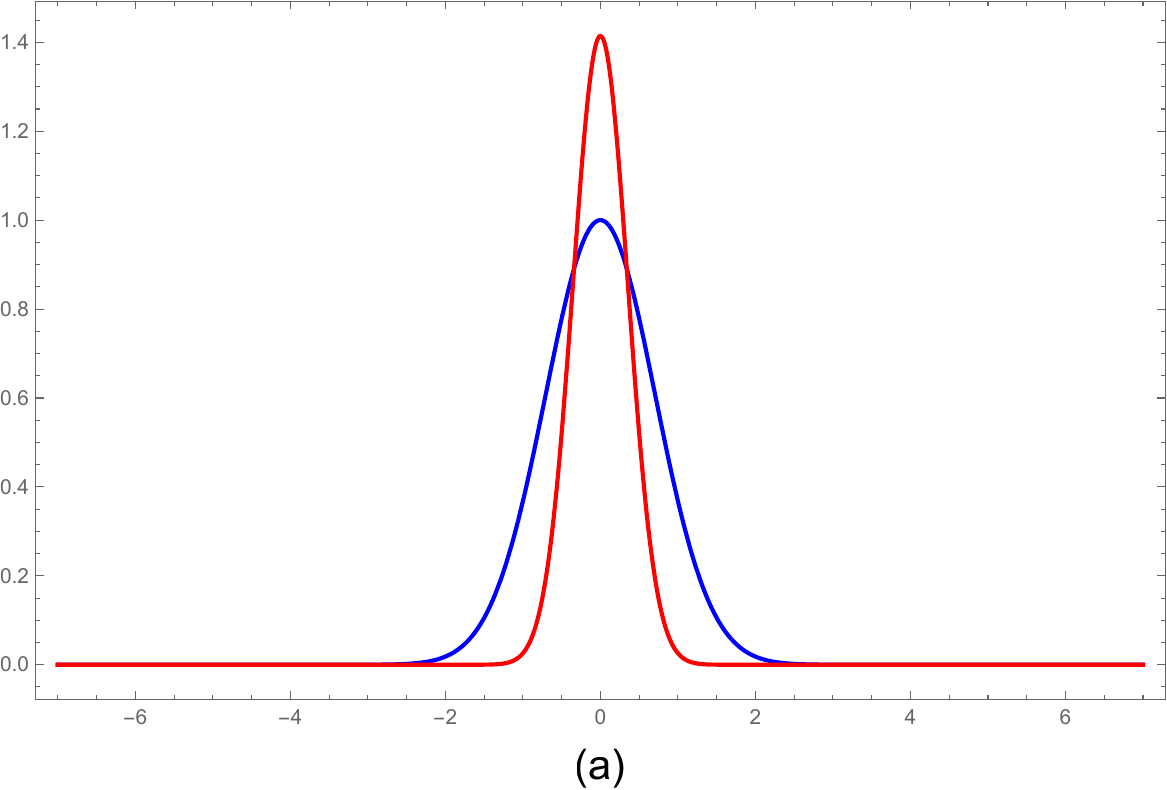}
    \includegraphics[scale=0.35]{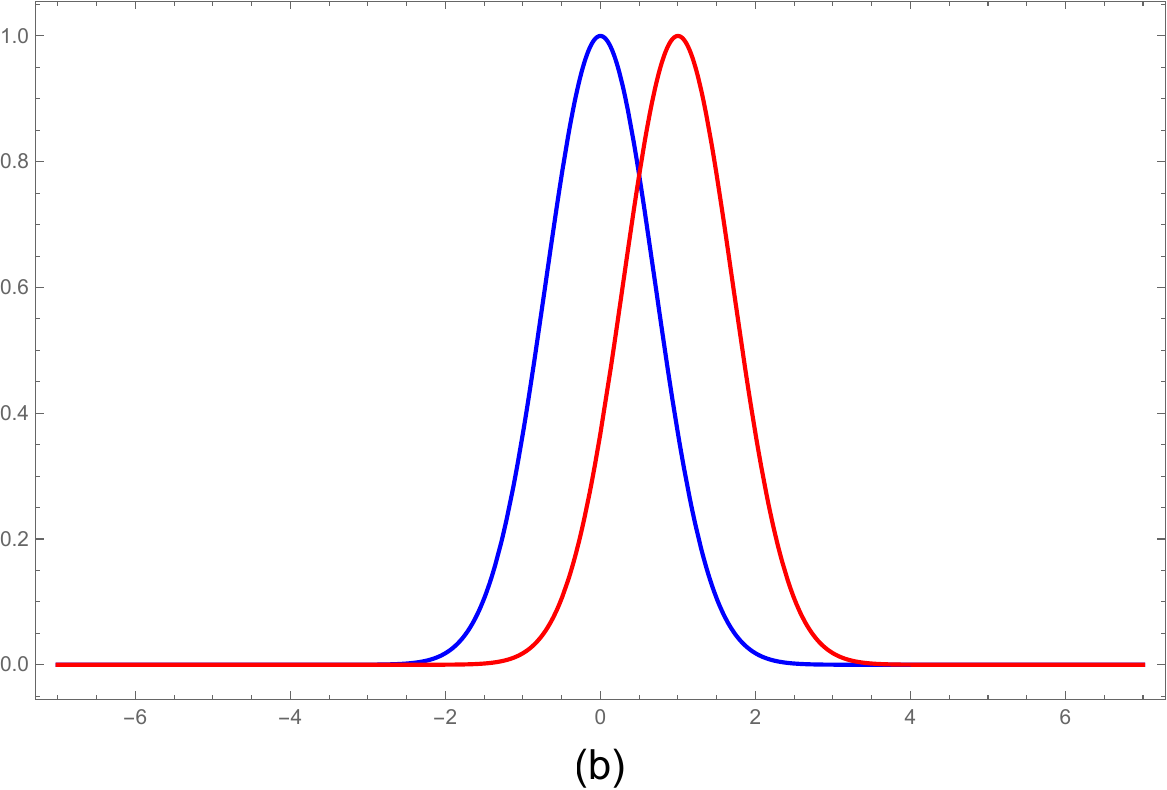}
\end{figure}

\begin{figure}
    \centering
    \caption{The red dotted line is the mother scaling function $s(x)$ for $K=3$, formed with the linear combination of the weighted sum of $6$ translated and scaled copies of itself.}
    \label{fig:scaling_equation_1}
    \includegraphics[scale=0.35]{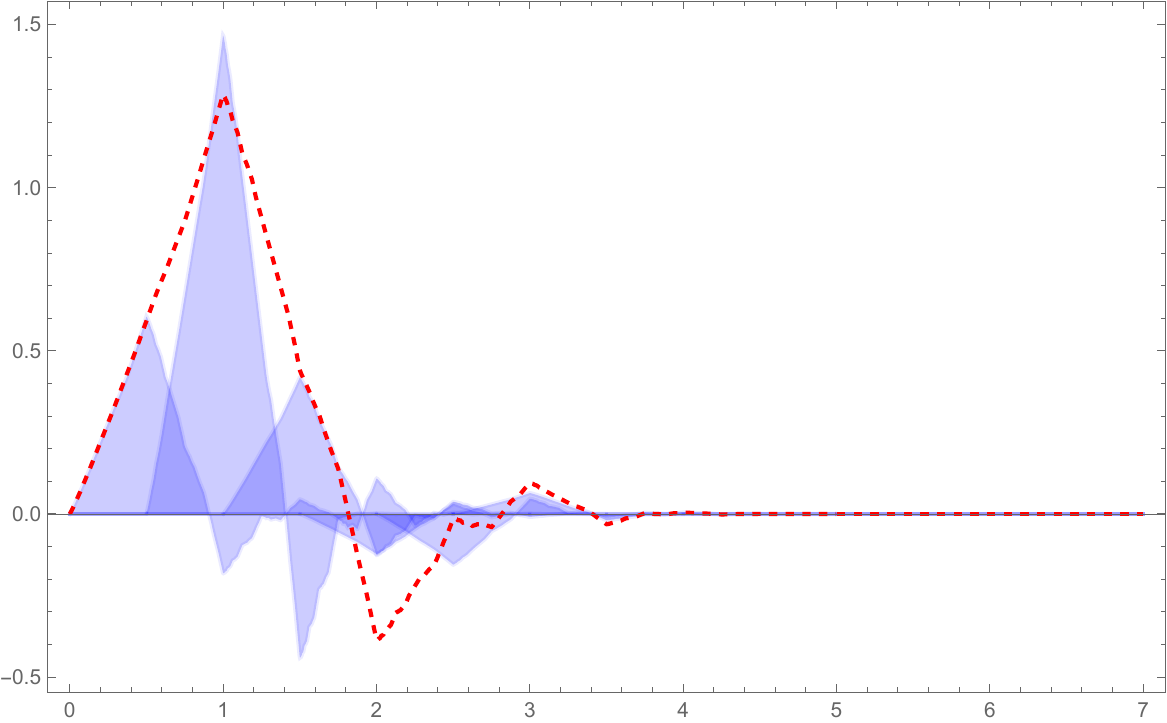}
\end{figure}

The elements of basis are constructed from a single function $s(x)$, called the mother scaling function, which is defined by the refinement equation,
\begin{eqnarray}
\label{eq:scaling_equation_2}
s(x)=\sum_{n=0}^{2K-1}h_n s(2x-n).
\end{eqnarray}
where $h_n$'s are called the filter coefficients. The refinement equation expresses the mother scaling function in terms of scaled and shifted versions of itself. 
This can be made explicit by defining the dyadic scaling operator, $\hat{D}$, and the translation operator, $\hat{T}$, 
\begin{eqnarray}
\hat{D}f(x)=\sqrt{2}f(2x),\quad \hat{T}g(x)=g(x-1),
\end{eqnarray}
where, $f(x)$ and $g(x)$, are arbitrary real functions. The operators, $\hat{D}$ and $\hat{T}$, are unitary with respect to the norm,
\begin{eqnarray}
\label{eq:norm}
(f,g) = \int f(x)g(x)dx
\end{eqnarray}
They do not commute with each other. Specifically, 
\begin{equation}
\label{noncommutative_D_and_T}
\hat{T}\hat{D} = \hat{D}\hat{T}^2.
\end{equation}
Fig. \ref{fig:D_and_T_operations} shows the graphical view of the action of $\hat{D}$ and $\hat{T}$. In terms of operators, $D$ and $T$, the form of the refinement equation,
\begin{eqnarray}
\label{eq:scaling_equation_1}
s(x)=\sum_{n=0}^{2K-1}{h_n}\hat{D}\hat{T}^ns(x),
\end{eqnarray}
shows that the mother scaling function of order $K$ is generated through a specific linear combination of $2K$ number of translated and dyadically scaled replicas of itself.  Eq. (\ref{eq:scaling_equation_1}) only determines the relative values of the mother scaling at different points on the real line. The absolute values of the mother scaling function at different points are fixed by demanding that 
\begin{eqnarray}
\label{eq:the_normalization_condition_of_scaling_function}
\int s(x)dx=1.
\end{eqnarray}
The procedure for determining the filter coefficients and the mother scaling function, $s(x)$ itself, starting from Eq. (\ref{eq:scaling_equation_1}) and Eq. (\ref{eq:the_normalization_condition_of_scaling_function}), is outlined in (cite references). In Fig. \ref{fig:scaling_equation_1}, we provide a visual representation of the mother scaling function and its associated refinement equation, Eq. (\ref{eq:scaling_equation_1}) for $K=3$.

The order $K$ determines the extent of smoothness and support of the mother scaling function. For example, the mother scaling function is singly differentiable for $K=2$ and doubly differentiable for $K=3$. Its support is given by the interval $[0,2K-1]$.  
\begin{figure}[hbt]
\centering
\caption{Order, $K=3$ Daubechies scaling functions for three different resolution $k=0,1$ and $2$ respectively.}
\label{fig:scaling_function_for_three_different_resolution}
\includegraphics[scale=0.35]{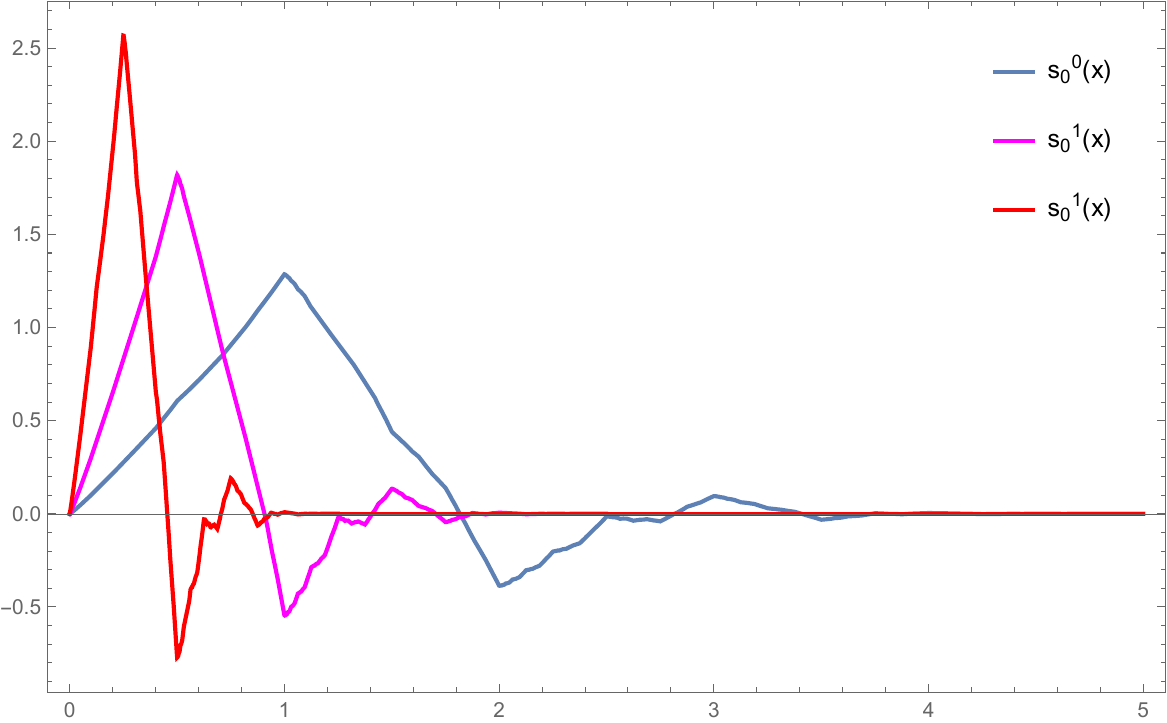}
\end{figure}

\begin{figure}[hbt]
\centering
\caption{Construction of the space of the square integrable functions by taking infinite resolution limit of the scaling function space $\mathcal{H}^k$.}
\label{fig:hilbert_space_only_scaling_function}
\includegraphics[scale=0.35]{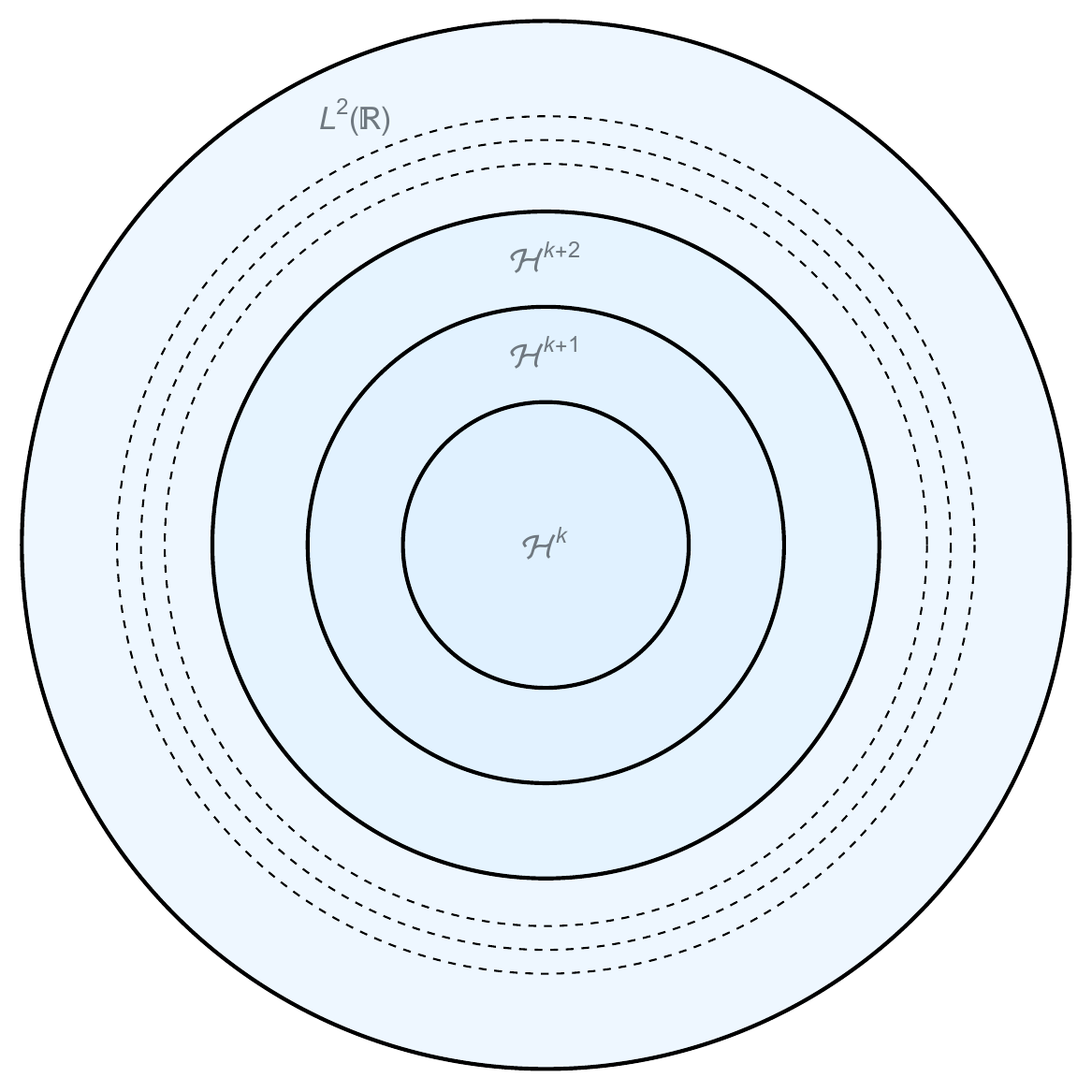}
\end{figure}

The resolution $k$ scaling basis functions are defined by,  
\begin{eqnarray}
\label{scaling_basis_functions}
s^k_n(x):=\hat{D}^k \hat{T}^n s(x),
\end{eqnarray}
with the location index $n$ ranging over the set of integers. Fig. \ref{fig:scaling_function_for_three_different_resolution} shows order $K=3$ Daubechies scaling functions for a few different resolutions and locations.
We demand that they constitute an orthonormal set, 
\begin{eqnarray}
\label{eq:orthonormality_of_the_scaling_function}
\int s^k_n(x)s^k_m(x)dx=\delta_{nm}.
\end{eqnarray}
The arbitrary linear combination of these functions generates the approximation space $\mathcal{H}^k$ of resolution $k$,
\begin{eqnarray}
\label{eq:formation_of_H_k}
\mathcal{H}^k=\left\{ f(x)|f(x)=\sum_{-\infty}^{\infty}f_n^{s,k} s^k_n(x),\sum_{n} |f^{s,k}_n|^2<\infty \right\}.
\end{eqnarray}
Using Eq. (\ref{noncommutative_D_and_T}), Eq. (\ref{eq:scaling_equation_1}) and Eq. (\ref{scaling_basis_functions}), we can show that resolution $k-1$ scaling basis functions can be expressed as linear combinations of resolution $k$ scaling basis functions,
\begin{eqnarray}
s^{k-1}_n(x)&=&\sum_{m=2n}^{2n+2K-1}H_{nm} s^k_{m}(x),
\end{eqnarray}
where $H_{nm} = h_{m-2n}$. This implies that the resolution $k-1$ approximation space is contained within resolution $k$ approximation space, 
\begin{eqnarray}
\mathcal{H}^{k-1}\subset \mathcal{H}^{k},
\end{eqnarray}
and by induction, for any $m>0$
\begin{eqnarray}
\mathcal{H}^k\subset \mathcal{H}^{k+m}.
\end{eqnarray}
Therefore, approximation spaces have a nested structure, as shown in Fig. \ref{fig:hilbert_space_only_scaling_function}. The infinite resolution limit of $\mathcal{H}^k$ generates the space of square-integrable functions $\mathcal{L}^2(\mathbb{R})$,
\begin{eqnarray}
\label{eq:the_scaling_function_representation_of_space_of_square_integrable_function}
\mathcal{L}^2(\mathbb{R})=\lim_{k\rightarrow \infty} \mathcal{H}^k.
\end{eqnarray}

\begin{figure}[hbt]
\centering
\caption{Order, $K=3$ Daubechies wavelet functions for different resolutions $k=0$, $1$ and $2$ respectively.}
\label{fig:wavelet_functions_for_different_resolutions}
\includegraphics[scale=0.35]{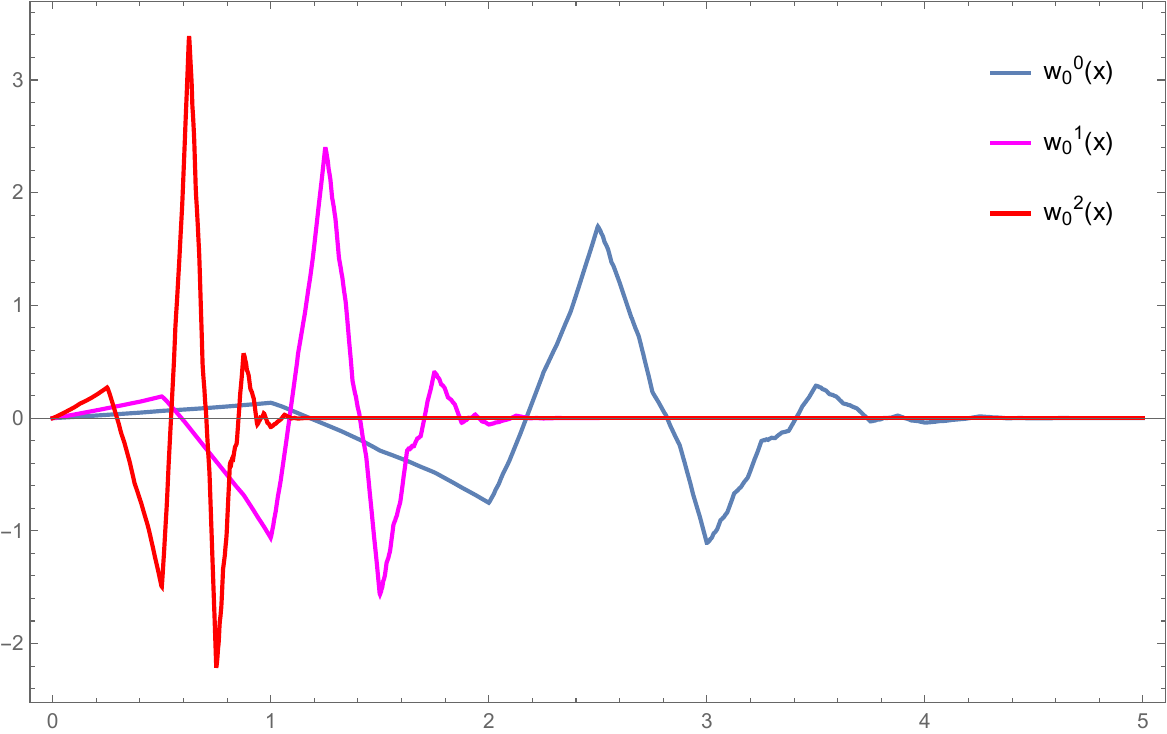}
\end{figure}

\begin{figure}[hbt]
\centering
\caption{Euler diagram of forming the space of square integrable functions ($\mathcal{L}^2(\mathbb{R})$) with scaling and wavelet basis.}
\label{fig:eular_diagram_of_forming_the_Hilbert_space_with_wavelet_basis}
\includegraphics[scale=0.35]{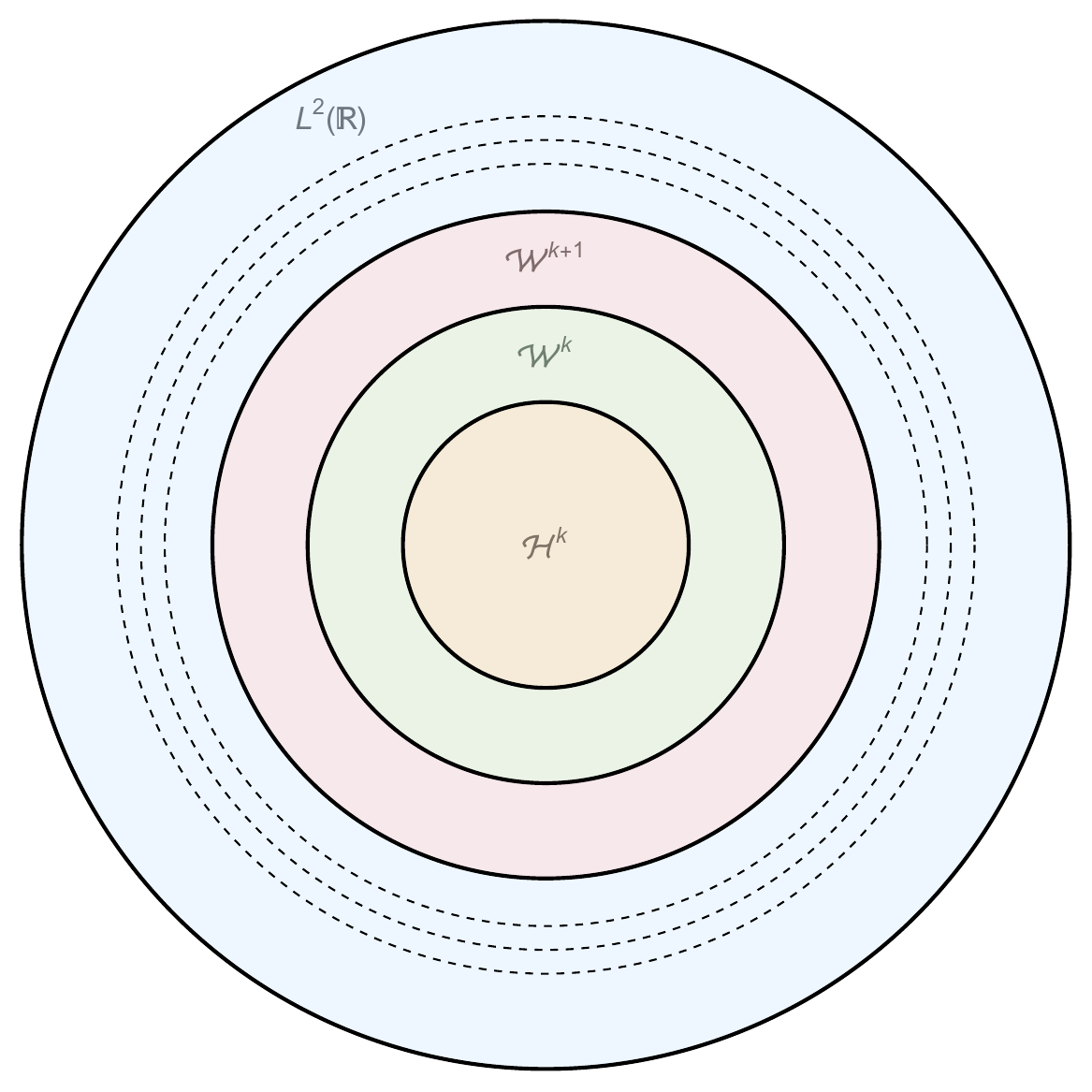}
\end{figure}

Since $\mathcal{H}^k$ is a proper subspace of $\mathcal{H}^{k+1}$, we can define $\mathcal{W}^{k}$ as the orthogonal complement of $\mathcal{H}^k$ in $\mathcal{H}^{k+1}$. $\mathcal{W}^{k}$ is referred to as the detail or wavelet space of resolution $k$. This implies that the resolution $k+1$ approximation space is a direct sum of resolution $k$, approximation and detail subspaces.
\begin{eqnarray}
\label{eq:space_H_(k+1)_by_scaling_and_wavelet_space}
\mathcal{H}^{k+1}=\mathcal{H}^k\oplus \mathcal{W}^k.
\end{eqnarray}
To construct the basis functions that span $\mathcal{W}^k$, we define the mother wavelet function, $w(x)$,  
\begin{eqnarray}
\label{eq:wavelet_equation_1}
w(x):=\sum_{n=0}^{2K-1} g_n \hat{D} \hat{T}^n s(x),
\end{eqnarray}
where the coefficients
\begin{eqnarray}
g_n:= (-1)^n h_{2K-1-n},
\end{eqnarray}
have been chosen such that the mother wavelet function $w(x)$ is orthogonal to $s(x)$. The basis functions for space $\mathcal{W}^k$, referred to as the resolution $k$ wavelet functions, are defined by,
\begin{eqnarray}
w^k_n(x):=\hat{D}^k\hat{T}^n w(x).
\end{eqnarray}
Fig. (\ref{fig:wavelet_functions_for_different_resolutions}) shows order $K=3$ Daubechies scaling functions for a few different resolutions and locations.
By design, the wavelet functions constitute an orthonormal set,
\begin{eqnarray}
\label{eq:orthonormality_of_wavelet_functions}
\int w^k_n(x)w^l_m(x)dx=\delta_{nm}\delta_{kl},
\end{eqnarray}
that are orthogonal to the scaling functions, 
\begin{eqnarray}
\label{eq:orthogonality_of_scaling_and_wavelet_functions}
\int s^k_m(x)w^{l+k}_n(x)dx=0, \quad l\geq 0,
\end{eqnarray}
The resolution $k$ wavelet functions span the resolution $k$ wavelet space $\mathcal{W}^k$,
\begin{eqnarray}
\mathcal{W}^k=\left\{f(x)|f(x)=\sum_{-\infty}^{\infty}f^{w,k}_n w^k_n(x),\sum_n 
|f^{w,k}_n|<\infty\right\}.
\end{eqnarray}
The space of square integrable functions, $\mathcal{L}^2(\mathbb{R})$, can be constructed by recursive use of Eq. (\ref{eq:space_H_(k+1)_by_scaling_and_wavelet_space}).
\begin{eqnarray}
\mathcal{L}^2(\mathbb{R})=\mathcal{H}^k\oplus \mathcal{W}^k\oplus \mathcal{W}^{k+1}\oplus \mathcal{W}^{k+2}....,
\end{eqnarray}
as depicted in Fig. \ref{fig:eular_diagram_of_forming_the_Hilbert_space_with_wavelet_basis}.

There are two possible basis choices for $\mathcal{H}^k$: one can either use the scaling functions $\left\{s^k_n(x)\right\}_{n=-\infty}^{\infty}$ of resolution $k$ or a combination of the scaling and wavelet functions of resolution $k-1$, $\left(\left\{s^{k-1}_n(x)\right\}_{n=-\infty}^{\infty}\cup \left\{w^{k-1}_n(x)\right\}_{n=-\infty}^{\infty}\right)$. These two bases are related to each other through an orthogonal transformation:
\begin{eqnarray}
s^{k-1}_n(x)&=&\sum_{m=2n}^{2n+2K-1}H_{nm} s^k_{m}(x),\\
w ^{k-1}_n(x)&=&\sum_{m=2n}^{2n+2K-1}G_{nm} s^k_{m}(x),\\
s^k_n(x)=\sum_{m=2n}^{2n+2K-1}&&H^t_{nm}s^{k-1}_m(x)+\sum_{m=2n}^{2n+2K-1}G^t_{nm}w^{k-1}_n(x),
\end{eqnarray}
where,
\begin{eqnarray}
H_{nm}=h_{m-2n},\quad G_{nm}=g_{m-2n}\\
H^t_{nm}=h_{n-2m},\quad G^t_{nm}=g_{n-2m}
\end{eqnarray}

The $h_n$ coefficients that are appearing in Eq. (\ref{eq:scaling_equation_1}), can be determined for any integral value of $K$ by solving the following system of equations:
\begin{eqnarray}
\label{eq:the_necessary_condition_to_find_out_h_n}
\sum_{n=0}^{2K-1}h_n &=&\sqrt{2},\\
\label{eq:the_orthonormal_condition_for_h_n}
\sum_{n=0}^{2K-1}h_nh_{n-2m}&=&\delta_{m0},\\
\label{eq:wavelet_functions_are_orthogonal_to_polynomial}
\sum_{n=0}^{2K-1}n^m g_n =\sum_{n=0}^{2K-1} &&n^m (-1)^n h_{2K-1-n},\nonumber\\
&=& 0, \quad m<K.
\end{eqnarray}
Equation (\ref{eq:the_necessary_condition_to_find_out_h_n}), can be found substituting the normalization condition of the scaling function, Eq (\ref{eq:the_normalization_condition_of_scaling_function}), into the scaling equation, Eq. (\ref{eq:scaling_equation_1}). Equation (\ref{eq:the_orthonormal_condition_for_h_n}), can be found using the scaling equation, Eq. (\ref{eq:scaling_equation_1}), and the orthonormality condition of the scaling functions, Eq. (\ref{eq:orthonormality_of_the_scaling_function}). Equation (\ref{eq:wavelet_functions_are_orthogonal_to_polynomial}), ensures that the wavelet functions are orthogonal to degree $K-1$ polynomials.

\begin{figure}[hbt]
\centering
\caption{The plot of scaling and wavelet function for different values of $K$.}
\label{fig:scaling_and_wavelet_functions_for_different_values_of_K}
\includegraphics[scale=0.30]{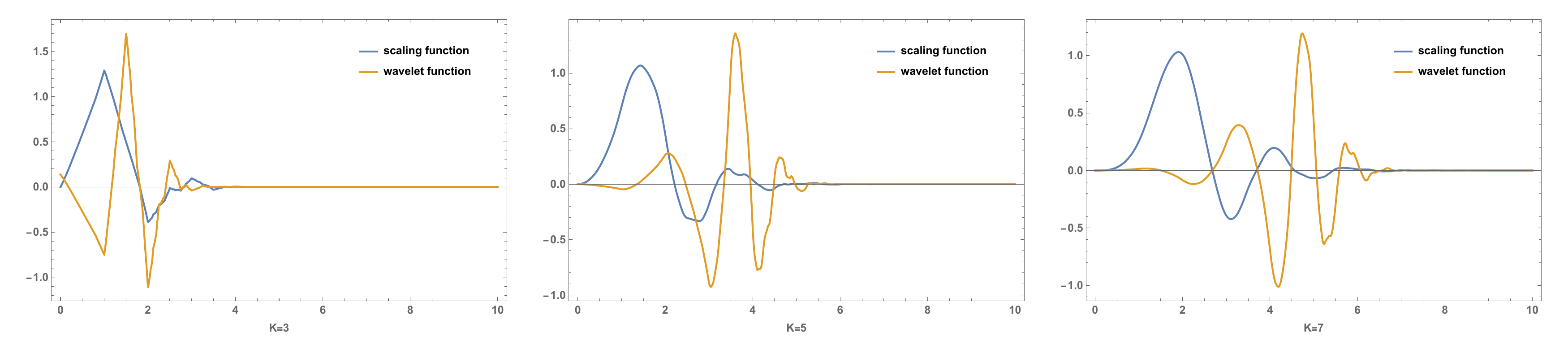}
\end{figure}

The $h_n$ coefficients for $K=1,2,3,4$ and $5$ are given in Table. \ref{tab:h_coefficient_of_Daubechies_wavelet_basis_for_different_K}.
\begin{table}[hbt]
\begin{center}
\setlength{\tabcolsep}{1.0pc}
\newlength{\digitwidth} \settowidth{\digitwidth}{\rm 0}
\catcode`?=\active \def?{\kern\digitwidth}
\caption{$h$ coefficients of Daubechies wavelet for different values of $K$.}
\label{tab:h_coefficient_of_Daubechies_wavelet_basis_for_different_K}
\vspace{1mm}
\begin{tabular}{c | c | c | c | c | c}
\specialrule{.15em}{.0em}{.15em}
\hline
$h_n$      & $K=1$        & $K=2$      & $K=3$        & $K=4$        & $K=5$ \\
\hline
$h_0$      & $0.707107$ & $0.482963$   & $0.332671  $ & $0.230378  $ & $0.160102$\\
$h_1$      & $0.707107$ & $0.836516$   & $0.806892  $ & $0.714847  $ & $0.603829$\\
$h_2$      & $0$        & $0.224144$   & $0.459878  $ & $0.630881  $ & $0.724309$\\
$h_3$      & $0$        & $-0.12941$   & $-0.135011 $ & $-0.0279838$ & $0.138428$\\
$h_4$      & $0$        & $0$          & $-0.0854413$ & $-0.187035 $ & $-0.242295$\\
$h_5$      & $0$        & $0$          & $0.0352263 $ & $0.0308414 $ & $-0.0322449$\\
$h_6$      & $0$        & $0$          & $0$          & $0.032883  $ & $0.0775715$\\
$h_7$      & $0$        & $0$          & $0$          & $-0.0105974$ & $-0.00624149$\\
$h_8$      & $0$        & $0$          & $0$          & $0$          & $-0.0125808$\\
$h_9$      & $0$        & $0$          & $0$          & $0$          & $3.33573\times 10^{-3}$\\
\hline
\specialrule{.15em}{.15em}{.0em}
\end{tabular}
\end{center}
\end{table}
Utilizing the values of $h_n$, we can find out the values of scaling function, $s(x)$ and the wavelet function, $w(x)$ for each points of $x$ recursively from Eq. (\ref{eq:scaling_equation_1}) and  Eq. (\ref{eq:wavelet_equation_1}). For any arbitrary value of $K$, the mother scaling function, $s(x)$ and the mother wavelet function, $w(x)$ will have the support within the interval $[0,2K-1]$. The function $s^k_n(x)$ and $w^k_n(x)$ will have the compact support smaller by $2^k$ in comparison with the $s(x)$
 and $w(x)$. The graphical view of the $s(x)$ and $w(x)$ for different values of $K=2,4$ and $6$ is given in Fig. \ref{fig:scaling_and_wavelet_functions_for_different_values_of_K}.
\begin{eqnarray}
&&s^k_n(x),w^k_n(x)\neq 0 \quad\forall x\in\left(\frac{n-0}{2^k},\frac{n+2K-1}{2^k}\right),\\
&&\implies \text{support size} =\frac{2K-1}{2^k}.
\end{eqnarray}

\section{Flow equations in Daubechies wavelet basis}
\label{sec:Flow-equation to separate scales and the coupling between two fields}

In this section, we outline the key elements of the flow-equation method introduced by Wegner \cite{https://doi.org/10.1002/andp.19945060203}. 

This method involves continuous evolution of the Hamiltonian $H(0)$ to its unitary equivalent form $H(\lambda)$. The unitary transformation $U(\lambda)$ is parametrized by a continuous parameter, $\lambda$. The flow equation method provides an alternative way to diagonalize or block-diagonalize the Hamiltonian using a unitary transformation. The transformed Hamiltonian will take the following form,
\begin{eqnarray}
\label{eq:h_lambda}
\mathrm{H}(\lambda)=U(\lambda)\mathrm{H}U^{\dagger}(\lambda).
\end{eqnarray} 
Here, $\mathrm{H}=\mathrm{H}(0)$, is the initial Hamiltonian. The unitary transformation, $U(\lambda)$, will satisfy the following differential equation,
\begin{eqnarray}
\frac{dU(\lambda)}{d\lambda}=\frac{dU(\lambda)}{d\lambda}U^{\dagger}U(\lambda)=K(\lambda)U(\lambda).
\end{eqnarray}
where, 
\begin{eqnarray}
\label{eq:the_generator}
K(\lambda)=\frac{dU(\lambda)}{d\lambda}U^{\dagger}(\lambda)=-K^{\dagger}(\lambda),
\end{eqnarray}
is the anti-hermitian generator of the unitary transformation. The chosen generator is used to transform the Hamiltonian into its desired unitary equivalent form as the value of $\lambda$ increases. 

It follows from the Eq. (\ref{eq:h_lambda}) and Eq. (\ref{eq:the_generator}), that the Hamiltonian, $\mathrm{H}(\lambda)$, will satisfy the following differential equation,
\begin{eqnarray}
\label{eq:flow-equation_of_hamiltonian}
\frac{d\mathrm{H}(\lambda)}{d\lambda}=\left[K(\lambda),\mathrm{H}(\lambda)\right].
\end{eqnarray}
Eq. (\ref{eq:flow-equation_of_hamiltonian}), is the flow-equation of the Hamiltonian which needs to be solved, with a suitable choice of the generator, to obtain a desired equivalent form of the Hamiltonian. 

\begin{figure}[H]
\begin{center}
\caption{The schematic diagram shows the volume and resolution truncated Hamiltonian divided into four components.}
\label{fig:Hamiltonian_divided_into_different_components}
\includegraphics[scale=0.35]{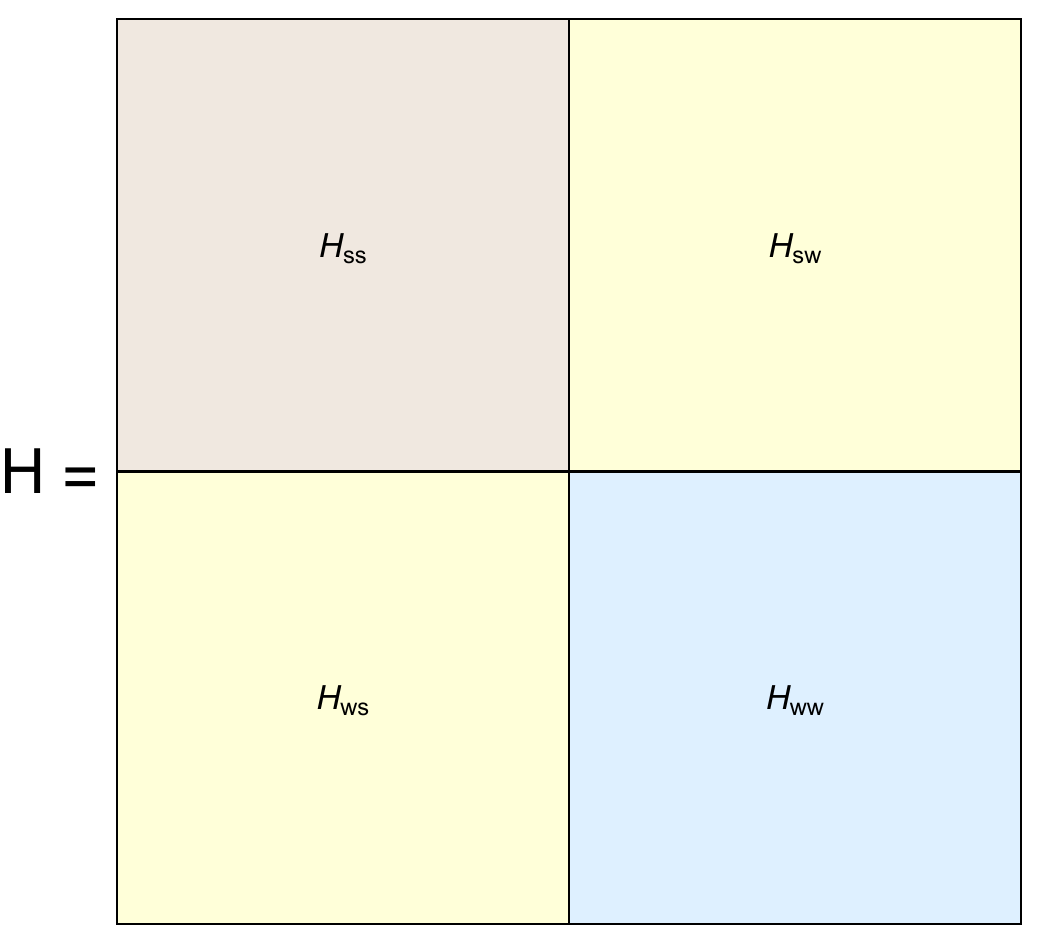}
\end{center}
\end{figure}

Now, we will present the solution of the flow-equation for the specific Hamiltonian matrix, represented in wavelet basis. To obtain a finite Hamiltonian matrix, we have to truncate the resolution of the basis functions and impose a restriction on the truncation index (truncation of volume). The schematic diagram for a truncated Hamiltonian to resolution $1$ with a specific finite volume is depicted in Fig. (\ref{fig:Hamiltonian_divided_into_different_components}).
The elements in the blocks $\mathrm{H}_{ss}$ and $\mathrm{H}_{ww}$ represent the coupling between degrees of freedom of the same scale. Whereas, the elements in the blocks $\mathrm{H}_{sw}$ and $\mathrm{H}_{ws}$ represent the coupling between different scale degrees of freedom.

\begin{figure}[hbt]
\caption{Schematic diagram of the generator $K(\lambda)$, which is constructed by taking the commutation of $G(\lambda)$ and $\mathrm{H}(\lambda)$, in wavelet basis.}
\label{fig:k_lambda}
\includegraphics[scale=0.35]{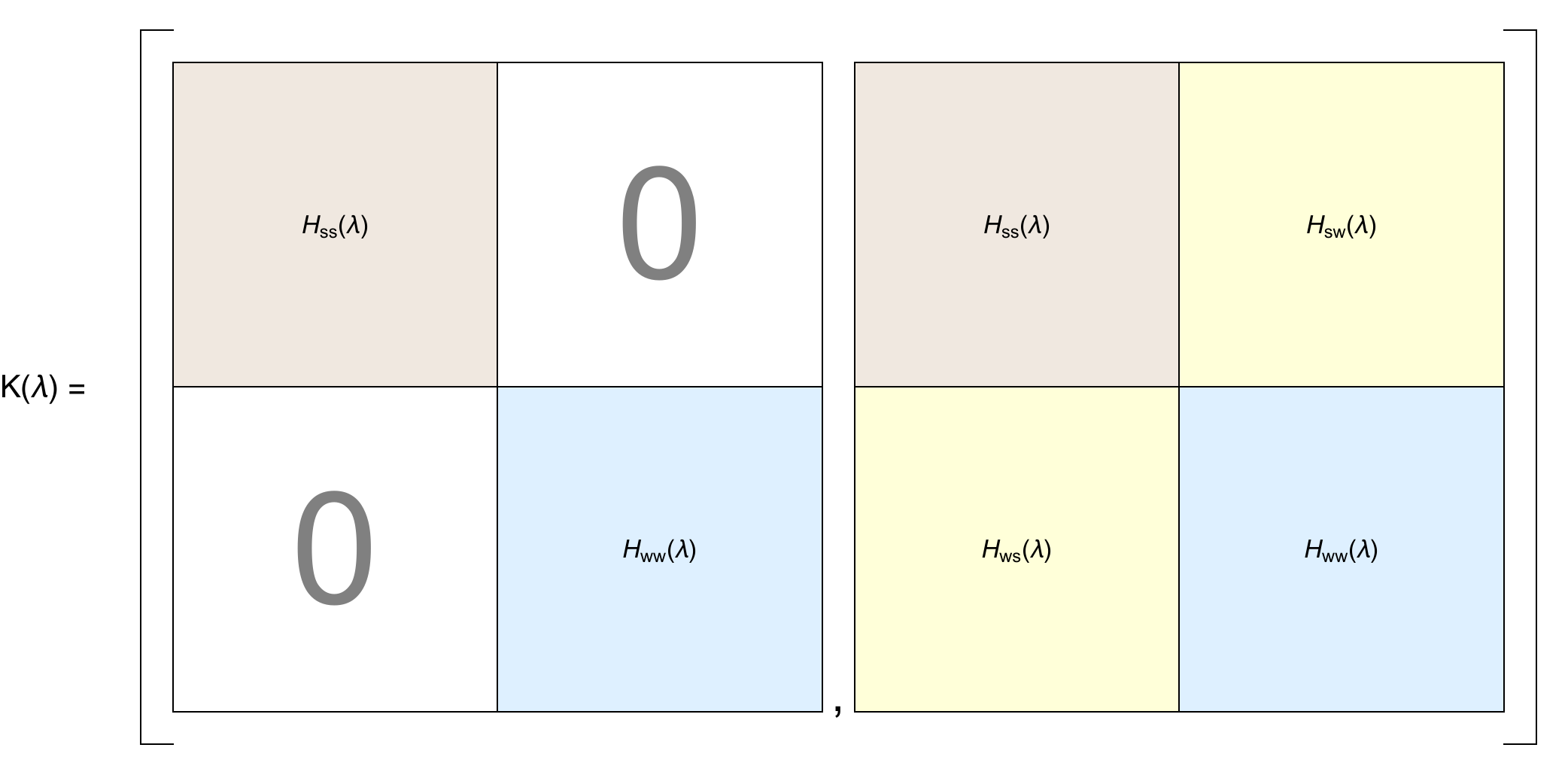},
\end{figure}

Using the flow equation, we want an derive an equivalent block diagonalized Hamiltonian of the initial Hamiltonian, where the same scale coupling terms will present and the different scale coupling term will be zero. To achieve this goal, we choose the generator, $K(\lambda)$, in following form,
\begin{eqnarray}
\label{eq:the_generator_choice_1}
K(\lambda)=\left[G(\lambda),\mathrm{H}(\lambda)\right],
\end{eqnarray}
where, $G(\lambda)$ is the part of the Hamiltonian, where different scale coupling terms do not present. The schematic diagram of $K(\lambda)$ is given in Fig. \ref{fig:k_lambda}.
As $G(\lambda)$ is Hermitian, the generator $K(\lambda)$ will be anti-Hermitian and Eq. (\ref{eq:flow-equation_of_hamiltonian}) will take the following form,
\begin{eqnarray}
\label{eq:the_flow_equation_2}
\frac{d\mathrm{H}(\lambda)}{d\lambda}=\left[\left[G(\lambda),\mathrm{H}(\lambda)\right],\mathrm{H}(\lambda)\right]=\left[\mathrm{H}(\lambda),\left[\mathrm{H}(\lambda),G(\lambda)\right]\right].
\end{eqnarray}

\begin{figure}[hbt]
\caption{The schematic diagram of total Hamiltonian, $\mathrm{H}(\lambda)$, at any arbitrary value of $\lambda$ as a sum of two matrices, $\mathrm{H}_b{\lambda}$ and $\mathrm{H}_c(\lambda)$.}
\label{fig:the_division_of_full_hamiltonian_hb_hc}
\includegraphics[scale=0.35]{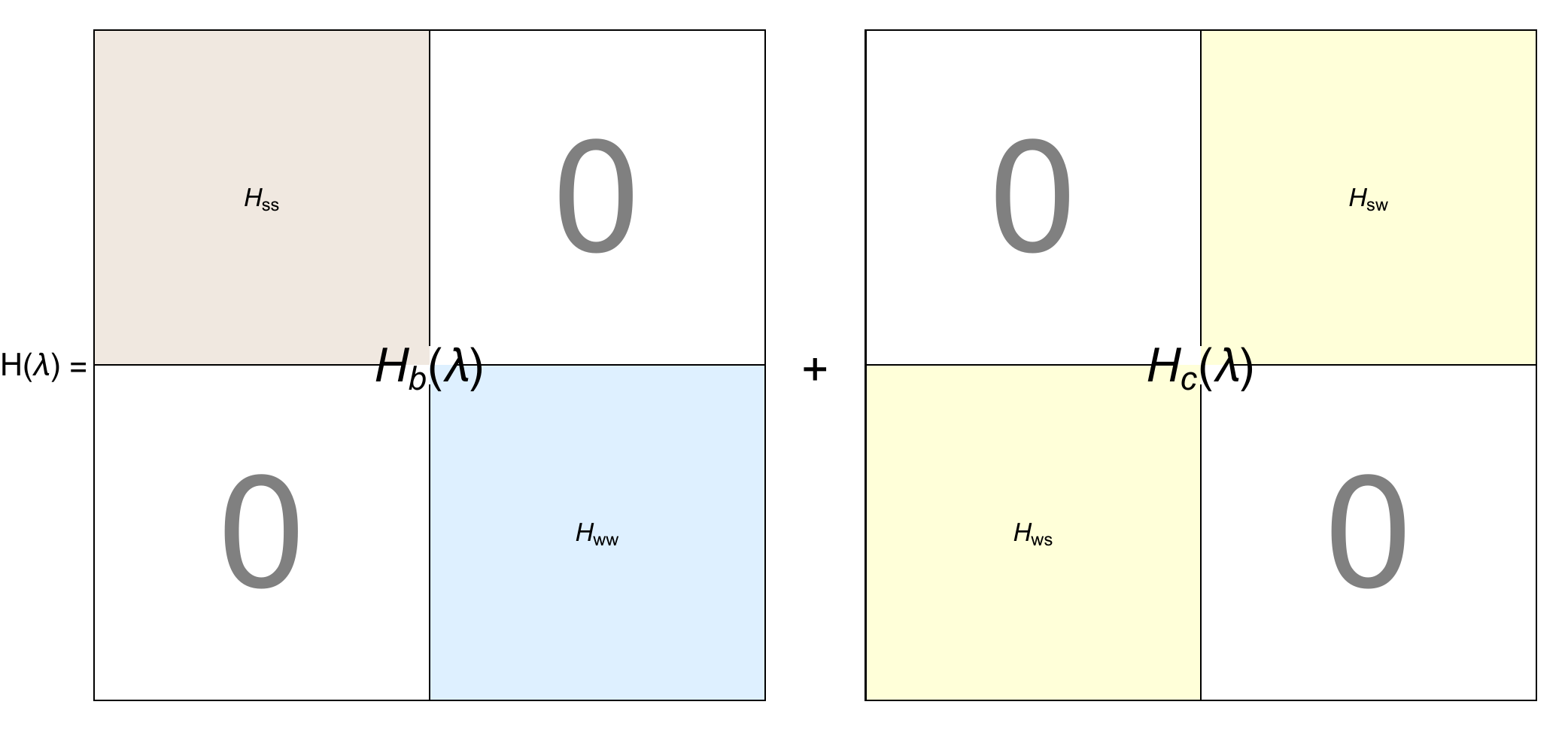}
\end{figure}

The full Hamiltonian, $\mathrm{H}(\lambda)$, for any arbitrary value of $\lambda$, Can be represented as a sum of two matrix, the same scale coupling matrix, $\mathrm{H}_b(\lambda)$ and the different scale coupling matrix, $\mathrm{H}_c(\lambda)$.
\begin{eqnarray}
\label{eq:the_total_hamiltonian}
\mathrm{H}(\lambda)=\mathrm{H}_b(\lambda)+\mathrm{H}_c(\lambda),
\end{eqnarray}
The schematic diagram of the decomposition of the full Hamiltonian, $\mathrm{H}(\lambda)$, is given in Fig. \ref{fig:the_division_of_full_hamiltonian_hb_hc}. This decomposition enables us to identify $G(\lambda)$ (the same scale coupling part of the Hamiltonian) to be $\mathrm{H}_b(\lambda)$.

The generator of the flow is given by,
\begin{eqnarray}
K(\lambda)=\left[\mathrm{H}_b(\lambda),\mathrm{H}_c(\lambda)\right]=\mathrm{H}_c(\lambda)',
\end{eqnarray}
where, $\mathrm{H}_c(\lambda)'$ is the different than that of the $\mathrm{H}_c(\lambda)$. So the right hand side of Eq. (\ref{eq:the_flow_equation_2}) will take the following form,
\begin{eqnarray}
\left[\mathrm{H}(\lambda),\mathrm{H}_c(\lambda)'\right]&=&\left[\mathrm{H}_c(\lambda)',\mathrm{H}_b(\lambda)+\mathrm{H}_c(\lambda)\right]\nonumber\\
&=& \left[\mathrm{H}_c(\lambda)',\mathrm{H}_b(\lambda)\right]+\left[H_c(\lambda)',\mathrm{H}_c(\lambda)\right]\nonumber\\
&=& \mathrm{H}_c(\lambda)''+\mathrm{H}_b(\lambda)'',
\end{eqnarray}
where, $\mathrm{H}_b(\lambda)''=\left[\mathrm{H}_b(\lambda),\mathrm{H}_c(\lambda)'\right]$ and $\mathrm{H}_c(\lambda)''=\left[\mathrm{H}_c(\lambda),\mathrm{H}_c(\lambda)'\right]$ are the same scale and different scale coupling terms of the right hand side of Eq. (\ref{eq:the_flow_equation_2}). Now, the flow equation can be separated into two parts the same scale coupling and the different scale coupling part of the Hamiltonian,
\begin{eqnarray}
\label{eq:flow_equation_3}
\frac{d\mathrm{H}_b(\lambda)}{d\lambda}=\mathrm{H}_b(\lambda)''=\left[\left[\mathrm{H}_b(\lambda),\mathrm{H}_c(\lambda)\right],\mathrm{H}_c(\lambda)\right],
\end{eqnarray}
and
\begin{eqnarray}
\label{eq:flow_equation_4}
\frac{d\mathrm{H}_c(\lambda)}{d\lambda}=\mathrm{H}_c(\lambda)''&=&\left[\left[\mathrm{H}_b(\lambda),\mathrm{H}_c(\lambda)\right],\mathrm{H}_b(\lambda)\right]\nonumber\\
&=& -\left[\left[\mathrm{H}_c(\lambda),\mathrm{H}_b(\lambda)\right],\mathrm{H}_b(\lambda)\right].
\end{eqnarray}
This equations have a symmetric form under the exchange of $\mathrm{H}_b(\lambda)\leftrightarrow \mathrm{H}_c(\lambda)$ except for the sign.

To understand how these equations will evolve the Hamiltonian to a desired form, we express the first equation in the basis of the eigenstates of $\mathrm{H}_c(\lambda)$ with eigenvalue $e_{cn}(\lambda)$ and express the second equation in the basis of the eigenstates of $\mathrm{H}_b(\lambda)$ with eigenvalue $e_{bn}(\lambda)$. 
\begin{eqnarray}
\frac{d\mathrm{H}_{bmn}(\lambda)}{d\lambda}=\left(e_{cm}(\lambda)-e_{cn}(\lambda)\right)^2 \mathrm{H}_{bmn}(\lambda),
\end{eqnarray}
and
\begin{eqnarray}
\frac{d\mathrm{H}_{cmn}(\lambda)}{d\lambda}=-\left(e_{bm}(\lambda)-e_{bn}(\lambda)\right)^2 \mathrm{H}_{cmn}(\lambda)
\end{eqnarray}
These equations can be solved exactly,
\begin{eqnarray}
\mathrm{H}_{bmn}(\lambda)&=& e^{\int_{0}^{\lambda}\left(e_{cm}(\lambda')-e_{cn}(\lambda')\right)^2 d\lambda'}\mathrm{H}_{bmn}(\lambda),\\
\mathrm{H}_{cmn}(\lambda)&=& e^{-\int_{0}^{\lambda}\left(e_{bm}(\lambda')-e_{bn}(\lambda')\right)^2 d\lambda'}\mathrm{H}_{cmn}(\lambda).
\end{eqnarray}
These solutions show that the matrix elements of $H_b$ increase exponentially, while the matrix elements of $H_c$ decrease exponentially with the increase of the flow parameter $\lambda$. This progression may halt if there are degeneracies in the eigenvalues, approximate degeneracies, or if the eigenvalues parametrized by $\lambda$ intersect.

It can be seen from these equations that when we are dealing with high resolution and large volume, the spectrum of the block diagonal operators will approach a continuous spectrum. In this case, we can expect to have closely spaced eigenvalues, which will result in slow convergence of certain parts of the scale-coupling operator.

In order to solve these equations numerically, the Hamiltonian needs to be truncated to a finite number of degrees of freedom. This requires truncating both the volume and the resolution. Any system with finite energy in a finite volume is expected to be dominated by a finite number of degrees of freedom \cite{doi:10.1137/1027082}. Degrees of freedom can be categorized into those associated with an experimental scale and other pertinent degrees of freedom at smaller scales. We can use scaling function fields as the degrees of freedom on the experimental scale and wavelet degrees of freedom on the smaller scales, which are still relevant to the given volume and energy scale.

When dealing with Hamiltonians with interactions, a different flow generator may be required to separate the desired degrees of freedom. It may also be necessary to first project the truncated Hamiltonian onto a subspace before solving the flow equation.

\section{The scalar field theory, the normal mode frequencies and the flow-equation in wavelet basis}
\label{sec:The_scalar_field_theory_and_the_normal_mode_frequencies}
The wavelet-based flow equations were first studied in the context of a single real free scalar field by Polyzou and Michlin \cite{PhysRevD.95.094501}. The study was carried out by truncating the theory to resolution $1$. The scale separation induced by the flow equations was demonstrated by computing the Hilbert-Schmidt norms so as to demonstrate the decoupling of the long- and short-distance variables. The normal mode frequencies of the effective $k=0$ resolution Hamiltonian generated by the flow equations were shown to contain effects of the $k=1$ resolution. In this section, we provide a short review of their work. We extend their work to one resolution higher and thereby demonstrate that the wavelet-based flow equations eliminate couplings across resolutions by driving the truncated Hamiltonian to block-diagonal form.

The Lagrangian density of the $1+1$-dimensional real scalar field theory is given by,
\begin{eqnarray}
\label{eq:free_scalar_field_theory_lagrangian}
\mathcal{L}(\phi(\mathrm{x},t),\partial_x \phi(\mathrm{x},t), \dot{\phi}(\mathrm{x},t))=\frac{1}{2}\left[\dot{\phi}^2(\mathrm{x},t)-(\partial_x \phi(\mathrm{x},t))^2-\mu^2 \phi(\mathrm{x},t)^2\right],
\end{eqnarray}
its corresponding Hamiltonian being,
\begin{eqnarray}
\label{eq:hamiltonian_scalar_field_theory}
\mathrm{H}(\phi(x,t),\pi(x,t))=\int dx \frac{1}{2}\left[\pi(x,t)^2+\left(\partial_x \phi(x,t)\right)^2+\mu^2 \phi(x,t)^2\right] ,
\end{eqnarray}
where, $\pi(x)$ represents the canonical momentum,
\begin{eqnarray}
\pi(x,t)=\frac{\partial \mathcal{L}}{\partial \dot{\phi}(x,t)}=\dot{\phi}(x,t).
\end{eqnarray}
The instant-form canonical quantization of the scalar field is carried out by demanding that $\phi(\mathrm{x},t)$ and $\pi(\mathrm{x},t)$ satisfy the equal-time canonical commutation relations,
\begin{eqnarray}
\label{eq:cannonical_commutation_relation_phi_pi}
\left[\phi(x,t),\pi(y,t)\right]&=&\iota \delta(x-y),\\
\label{eq:connonical_commutation_relation_phi_phi_pi_pi}
\left[\pi(x,t),\pi(y,t)\right]&=&\left[\phi(x,t),\phi(y,t)\right]=0.
\end{eqnarray}
We assume that the scalar field exists only within an interval of $0\leq x \leq L$ with hard boundaries. We write $\phi(x,t)$ and $\pi(x,t)$ in terms of the normal modes as follows,
\begin{eqnarray}
\label{eq:phi_xt_interms_of_normal_modes}
\phi(x,t)&=&\sum_{p=1}^{\infty} \phi_p(t) \sqrt{\frac{2}{L}}\sin(\frac{p\pi x}{L}),\\
\label{eq:pi_xt_in_terms_of_normal_modes}
\pi(x,t)&=&\sum_{p=1}^{\infty} \pi_p(t) \sqrt{\frac{2}{L}}\sin(\frac{p\pi x}{L}).
\end{eqnarray}
On substituting Eq. (\ref{eq:phi_xt_interms_of_normal_modes}) and Eq. (\ref{eq:pi_xt_in_terms_of_normal_modes}) into Eq. (\ref{eq:free_scalar_field_theory_lagrangian}), we get,
\begin{eqnarray}
\mathrm{H}=\sum_{p=1}^{\infty}\frac{1}{2} \left(\pi_p^2(t)+\omega_p^2\phi_p^2(t)\right),
\end{eqnarray}
where, $\omega_p = \sqrt{\left(\frac{p\pi}{L}\right)^2+\mu^2}$, is the normal mode frequency corresponding to the $p$-mode. The free scalar field is thus represented by a set of uncoupled simple harmonic oscillators with normal mode frequencies given by $\omega_p$.

The wavelet-based representation of Hamiltonian dynamics is constructed by resolving $\phi(x)$ and $\pi(x)$ in the discrete Daubechies wavelet basis:
\begin{eqnarray}
\label{eq:expansion_of_scalar_field_in_wavelet_basis}
\phi(\mathrm{x},t)&=&\sum_{n=-\infty}^{\infty}\phi^{s,k}_n(t)s^k_n(x)+\sum_{l\geq k}^{\infty}\sum_{n=-\infty}^{\infty} \phi^{w,l}_n(t) w^l_n(x),\\
\label{eq:expansion_of_scalar_field_momentum_in_wavelet_basis}
\pi(\mathrm{x},t)&=&\sum_{n=-\infty}^{\infty}\pi^{s,k}_n(t)s^k_n(x)+\sum_{l\geq k}^{\infty}\sum_{n=-\infty}^{\infty} \pi^{w,l}_n(t) w^l_n(x),
\end{eqnarray}
where, the scaling coefficients $\phi^{s,k}_n$ and $\pi^{s,k}_n$, represent variables that describe physics of the scalar field down to the length scale $\frac{2K-1}{2^k}$. The physics of the scalar field on length scales finer than $\frac{2K-1}{2^k}$ is contained in the variables given by the wavelet coefficients $\{\phi^{w,l}_n, \pi^{w,l}_n \,|\, l\ge k \}$. 

Within the wavelet-based formulation, the quantization of the classical field theory is carried out in terms of the variables given by the scaling and wavelet coefficients. Substituting Eq. (\ref{eq:expansion_of_scalar_field_in_wavelet_basis}) and Eq. (\ref{eq:expansion_of_scalar_field_momentum_in_wavelet_basis}) into Eq. (\ref{eq:hamiltonian_scalar_field_theory}) and using the orthogonality of the wavelet basis, Eq. (\ref{eq:orthonormality_of_the_scaling_function}), Eq. (\ref{eq:orthonormality_of_wavelet_functions}) and Eq. (\ref{eq:orthogonality_of_scaling_and_wavelet_functions}), gives the Hamiltonian in terms of the scaling and the wavelet variables.
\begin{eqnarray}
\mathrm{H}=\mathrm{H}_{ss}+\mathrm{H}_{ww}+\mathrm{H}_{sw}
\end{eqnarray}
where $\mathrm{H}_{ss}$ is part of the Hamiltonian that describes physics from the coarsest length scale down to the length scale $\frac{2K-1}{2^k}$.
\begin{eqnarray}
\label{eq:hss_scalar_field_theory}
\mathrm{H}_{ss}:=\frac{1}{2}\left(\sum_n \pi^{s,k}_n(t)\pi^{s,k}_n(t)+\sum_n \phi^{s,k}_m(t)\phi^{s,k}_n(t) \mathcal{D}^k_{ss,mn}+\sum_n \mu^2 \phi^{s,k}_n(t) \phi^{s,k}_n(t)\right).
\end{eqnarray}
 In effect, $\mathrm{H}_{ss}$ represents the Hamiltonian truncated to resolution $k$. Similarly, $\mathrm{H}_{ww}$ is part of the Hamiltonian that describes physics on all the length scales finer than $\frac{2K-1}{2^k}$,
\begin{eqnarray}
\label{eq:hww_scalar_field_theory}
\mathrm{H}_{ww}:=\frac{1}{2}\left(\sum_n \pi^{w,l}_n(t)\pi^{w,l}_n(t)+\sum_n \phi^{w,l}_m(t)\phi^{w,q}_n(t) \mathcal{D}^{lq}_{ww,mn}+\sum_n \mu^2 \phi^{w,l}_n(t) \phi^{w,l}_n(t)\right).
\end{eqnarray}
The coupling between the length scales coarser and finer than $\frac{2K-1}{2^k}$ is given by,
\begin{eqnarray}
\label{eq:hsw_scalar_field_theory}
\mathrm{H}_{sw}&:=&\frac{1}{2}\left(\sum_{m,q,n}\phi^{s,k}_m(t)\phi^{w,q}_n(t) \mathcal{D}^{kq}_{sw,mn}\right).
\end{eqnarray}
In Eq. (\ref{eq:hss_scalar_field_theory}), Eq. (\ref{eq:hww_scalar_field_theory}), and Eq. (\ref{eq:hsw_scalar_field_theory}), $\mathcal{D}^k_{ss,mn}$, $\mathcal{D}^{lq}_{ww,mn}$, $\mathcal{D}^{kq}_{sw,mn}$ denote the overlap integrals,
\begin{eqnarray}
\mathcal{D}^{k}_{ss,mn}&=& \int dx \frac{\partial s^k_m(x)}{\partial x}\frac{\partial s^k_n(x)}{\partial x},\\
\mathcal{D}^{lq}_{ww,mn}&=& \int dx \frac{\partial w^l_m(x)}{\partial x}\frac{\partial w^q_n(x)}{\partial x},\\
\mathcal{D}^{kq}_{sw,mn}&=& \int dx \frac{\partial s^k_m(x)}{\partial x}\frac{\partial w^q_n(x)}{\partial x}.
\end{eqnarray}
The procedure to evaluate these overlap integrals analytically is provided in \cite{PhysRevD.87.116011}.

The inverse of Eq. (\ref{eq:expansion_of_scalar_field_in_wavelet_basis}) and Eq. (\ref{eq:expansion_of_scalar_field_momentum_in_wavelet_basis}) (mathematically, the discrete wavelet transform (DWT)), 
\begin{eqnarray}
\phi^{s,k}_n=\int \phi(x)s^k_n(x)dx,\quad \phi^{w,l}_n=\int \phi(x) w^l_n(x)dx,\\
\pi^{s,k}_n=\int \pi(x)s^k_n(x)dx,\quad \pi^{w,l}_n=\int \pi(x) w^l_n(x)dx,
\end{eqnarray}
and the equal time commutation relations, Eq. (\ref{eq:cannonical_commutation_relation_phi_pi}) and Eq. (\ref{eq:connonical_commutation_relation_phi_phi_pi_pi}), lead to the canonical commutation relations between the scaling variables, $\{\phi^{s,k}_{n},\pi^{s,k}_n\}$ and the wavelet variables, $\{\phi^{w,l}_{n},\pi^{w,l}_n\}$,
\begin{eqnarray}
\left[\phi^{s,k}_m(t),\pi^{s,k}_n(t)\right]&=&\iota \delta_{mn},\quad \left[\phi^{s,k}_m(t),\phi^{s,k}_n(t)\right]=0, \quad \left[\pi^{s,k}_m(t),\pi^{s,k}_n(t)\right]=0,\\
\left[\phi^{w,l}_m(t),\pi^{w,q}_n(t)\right]&=&\iota \delta_{lq}\delta_{mn},\quad \left[\phi^{w,l}_m(t),\phi^{w,q}_n(t)\right]=0, \quad \left[\pi^{w,l}_m(t),\pi^{w,q}_n(t)\right]=0,\\
&&\left[\phi^{s,k}_m(t),\phi^{w,l}_n(t)\right]=0,\quad \left[\pi^{s,k}_m(t),\pi^{w,l}_n(t)\right]=0,\\
&&\left[\phi^{s,k}_m(t),\pi^{w,l}_n(t)\right]=0, \quad \left[\phi^{w,l}_m(t),\pi^{s,k}_n(t)\right]=0.
\end{eqnarray}

In terms of these variables, the full Hamiltonian can be rewritten in a compact matrix form as,
\begin{eqnarray}
\label{eq:compact_full_Hamiltonian_scalar_field_theory}
\mathrm{H}=\sum_{m,n=-\infty}^{\infty}\frac{1}{2}\left[
\begin{pmatrix}
\pi^{s,k}_m & \pi^{w,l}_m
\end{pmatrix}
\begin{pmatrix}
\delta_{mn} & 0 \\
0 & \delta_{mn}
\end{pmatrix}
\begin{pmatrix}
\pi^{s,k}_n\\
\pi^{w,l}_n
\end{pmatrix}
+
\begin{pmatrix}
\phi^{s,k}_m & \phi^{w,l}_m
\end{pmatrix}
\begin{pmatrix}
\mu^2\delta_{mn}+\mathcal{D}^{k}_{ss,mn} & \mathcal{D}^{kq}_{sw,mn}\\
\mathcal{D}^{qk}_{ws,mn} & \mu^2\delta_{mn}+\mathcal{D}^{lq}_{ww,mn}
\end{pmatrix}
\begin{pmatrix}
\phi^{s,k}_n\\
\phi^{w,l}_n
\end{pmatrix}
\right].\quad
\end{eqnarray}
Eq. (\ref{eq:compact_full_Hamiltonian_scalar_field_theory}) shows that within the wavelet formulation, the free scalar field theory is represented by a set of linearly coupled harmonic oscillators.
The coupling matrix,
\begin{eqnarray}
\label{eq:omega_square_matrix_scalar_field_theory}
\Omega^2:= \begin{pmatrix}
\mu^2\delta_{mn}+\mathcal{D}^{k}_{ss,mn} & \mathcal{D}^{kq}_{sw,mn}\\
\mathcal{D}^{qk}_{ws,mn} & \mu^2\delta_{mn}+\mathcal{D}^{lq}_{ww,mn}
\end{pmatrix}
\end{eqnarray}
is a real symmetric matrix, which can always be diagonalized using a similarity transformation. There exists an orthogonal matrix $O$ such that
\begin{eqnarray}
O^T \Omega^2 O:=
\begin{pmatrix}
{\omega_s}^2 & 0\\
0 & {\omega_w}^2
\end{pmatrix}.
\end{eqnarray}
where,
${\omega_s}^2$ and ${\omega_w}^2$ are the diagonal matrices containing the eigenvalues of matrix $\Omega^2$. These eigenvalues represent the squares of the normal-mode frequencies. The orthogonal matrix can be used to construct the normal mode variables in terms of the scaling and wavelet variables.

The comparison of the exact values of the squares of the normal mode frequencies $\omega_p^2 $ with those computed within the wavelet formulation for increasing resolution within the spatial interval $L=20$, mass $\mu =1$ is presented in Table \ref{tab:normal_mode_frequencies_exact_and_for_res_10_scalar_fields}. The calculation of eigenvalues of $\Omega^2$, has been done using the scaling function representation of the theory. The multi-scale (scaling-wavelet function) representation is used to separate the coupling between two scales using the flow equation method and to obtain an equivalent Hamiltonian consisting of only physically relevant degrees of freedom. 
\begin{table}[H]
\begin{center}
\setlength{\tabcolsep}{0.5pc}
\catcode`?=\active \def?{\kern\digitwidth}
\caption{Comparison of the exact square of the normal mode frequencies with the square of the normal mode frequencies computed using the wavelet techniques for different values of resolution ($k$)}
\label{tab:normal_mode_frequencies_exact_and_for_res_10_scalar_fields}
\vspace{1mm}
\begin{tabular}{c | c | c | c | c}
\specialrule{.15em}{.0em}{.15em}
\hline
Exact $\omega_p^2$ & $\omega_p^2$ for $k=0$ & $\omega_p^2$ for $k=1$ & $\omega_p^2$ for $k=6$ & $\omega_p^2$ for $k=10$\\
\hline
$1.024674$ & $1.036206$ & $1.029601$ & $1.024809$ & $1.024682$\\
$1.098696$ & $1.145666$ & $1.118457$ & $1.099235$ & $1.098730$\\
$1.222066$ & $1.332554$ & $1.266802$ & $1.223278$ & $1.222142$\\
$1.394784$ & $1.608288$ & $1.475262$ & $1.396938$ & $1.394918$\\
$1.616850$ & $1.994948$ & $1.745183$ & $1.620216$ & $1.617060$\\
$1.888264$ & $2.525095$ & $2.079026$ & $1.893112$ & $1.888566$\\
$2.209027$ & $3.236390$ & $2.480787$ & $2.215624$ & $2.209437$\\
$2.579137$ & $4.161702$ & $2.956357$ & $2.587754$ & $2.579673$\\
$2.998595$ & $5.316987$ & $3.513791$ & $3.009502$ & $2.999274$\\
$3.467401$ & $6.689941$ & $4.163396$ & $3.480866$ & $3.468239$\\
$3.985555$ & $8.232543$ & $4.917626$ & $4.001848$ & $3.986570$\\
$4.553058$ & $9.859974$ & $5.790746$ & $4.572448$ & $4.554265$\\
$5.169908$ & $11.45703$ & $6.798267$ & $5.192665$ & $5.171325$\\
$5.836106$ & $12.89138$ & $7.956184$ & $5.862500$ & $5.837750$\\
$6.551652$ & $14.03144$ & $9.280038$ & $6.581952$ & $6.553539$\\
$7.316547$ & $14.76557$ & $10.78385$ & $7.351022$ & $7.318693$\\
\hline
\specialrule{.15em}{.15em}{.0em}
\end{tabular}
\end{center}
\end{table}

The scale separation, within the context of $(1+1)$-dimensional scalar field theory, using wavelet-based flow equations was first demonstrated by Michlin and Polyzou \cite{PhysRevD.95.094501}. This computation was restricted to resolution $k=1$. We present an extension to this computation to one resolution higher. Through this computation, we demonstrate that the Hamiltonian flows to a block diagonal form, with each block labelled by resolution. To organize our computation, we rewrite the Hamiltonian, Eq. (\ref{eq:omega_square_matrix_scalar_field_theory}), in a matrix form using Eq. (\ref{eq:compact_full_Hamiltonian_scalar_field_theory}),
\begin{eqnarray}
\label{eq:compact_full_Hamiltonian_scalar_field_theory_1}
\mathrm{H}=\sum_{m,n=-\infty}^{\infty}\frac{1}{2}\left[
\begin{pmatrix}
\pi^{s,k}_m & \pi^{w,l}_m
\end{pmatrix}
\begin{pmatrix}
\delta_{mn} & 0 \\
0 & \delta_{mn}
\end{pmatrix}
\begin{pmatrix}
\pi^{s,k}_n\\
\pi^{w,l}_n
\end{pmatrix}
+
\begin{pmatrix}
\phi^{s,k}_m & \phi^{w,l}_m
\end{pmatrix}
\Omega^2
\begin{pmatrix}
\phi^{s,k}_n\\
\phi^{w,l}_n
\end{pmatrix}
\right].
\end{eqnarray}
As described in Section \ref{sec:Flow-equation to separate scales and the coupling between two fields}, the Hamiltonian is transformed to an equivalent form through a unitary transformation, parameterized by the continuous parameter $\lambda$,
\begin{eqnarray}
\mathrm{H}(\lambda)=U(\lambda) \mathrm{H} U^{\dagger}(\lambda)=\sum_{m,n=-\infty}^{\infty}\frac{1}{2}\left[
U(\lambda)
\begin{pmatrix}
\pi^{s,k}_m & \pi^{w,l}_m
\end{pmatrix}
\begin{pmatrix}
\pi^{s,k}_n\\
\pi^{w,l}_n
\end{pmatrix}
U^{\dagger}(\lambda)
+
U(\lambda)
\begin{pmatrix}
\phi^{s,k}_m & \phi^{w,l}_m
\end{pmatrix}
\Omega^2
\begin{pmatrix}
\phi^{s,k}_n\\
\phi^{w,l}_n
\end{pmatrix}
U^{\dagger}(\lambda)
\right].
\end{eqnarray}
We assume that this unitary transformation generates an orthogonal transformation $O$ of both the fields and their conjugate momenta. The unitary transformation representing the flow does not induce mixing between fields and their conjugate momenta. This is justified based on the results of \cite{PhysRevD.95.094501}.  Eq. (\ref{eq:compact_full_Hamiltonian_scalar_field_theory_1}) can then be rewritten as,
\begin{eqnarray}
\mathrm{H}(\lambda)=\sum_{m,n=-\infty}^{\infty}\frac{1}{2}\left[
\begin{pmatrix}
\pi^{s,k}_m & \pi^{w,l}_m
\end{pmatrix}
\begin{pmatrix}
\pi^{s,k}_n\\
\pi^{w,l}_n
\end{pmatrix}
+
\begin{pmatrix}
\phi^{s,k}_m & \phi^{w,l}_m
\end{pmatrix}
O^{T}(\lambda)
\Omega^2
O(\lambda)
\begin{pmatrix}
\phi^{s,k}_n\\
\phi^{w,l}_n
\end{pmatrix}
\right].
\end{eqnarray}
Differentiating both side with $\lambda$, we get,
\begin{eqnarray}
\frac{d\mathrm{H}(\lambda)}{d\lambda}&=&\sum_{m,n=-\infty}^{\infty}\frac{1}{2}\left[
\begin{pmatrix}
\phi^{s,k}_m & \phi^{w,l}_m
\end{pmatrix}\left(
\frac{dO^{T}(\lambda)}{d\lambda}O(\lambda)O^{T}(\lambda)
\Omega^2
O(\lambda)+
O^{T}(\lambda)
\Omega^2 O(\lambda)O^{T}(\lambda)
\frac{dO(\lambda)}{d\lambda}
\right)
\begin{pmatrix}
\phi^{s,k}_n\\
\phi^{w,l}_n
\end{pmatrix}
\right]\nonumber\\
\label{eq:modified_flow_equation}
&=& \sum_{m,n=-\infty}^{\infty}\frac{1}{2}\left[
\begin{pmatrix}
\phi^{s,k}_m & \phi^{w,l}_m
\end{pmatrix}\left(
\frac{dO^{T}(\lambda)}{d\lambda}O(\lambda)
\Omega^2(\lambda)-
\Omega^2 (\lambda)
\frac{dO^{T}(\lambda)}{d\lambda}O(\lambda)
\right)
\begin{pmatrix}
\phi^{s,k}_n\\
\phi^{w,l}_n
\end{pmatrix}
\right],
\end{eqnarray}
here, $\Omega^2(\lambda)=O^T(\lambda)\Omega^2 O(\lambda)$ and Eq. (\ref{eq:modified_flow_equation}), will take the following form,
\begin{eqnarray}
\frac{d\mathrm{H}(\lambda)}{d\lambda}= \sum_{m,n=-\infty}^{\infty}\frac{1}{2}\left[
\begin{pmatrix}
\phi^{s,k}_m & \phi^{w,l}_m
\end{pmatrix}
\left[K(\lambda),\Omega^2(\lambda)\right]
\begin{pmatrix}
\phi^{s,k}_n\\
\phi^{w,l}_n
\end{pmatrix}
\right],
\end{eqnarray}
where, $K(\lambda)$ is the generator of the transformation that satisfies,
\begin{eqnarray}
\label{eq:the_generator_2}
K(\lambda)=\frac{dO^{T}(\lambda)}{d\lambda}O(\lambda)=-O^{T}(\lambda)\frac{dO(\lambda)}{d\lambda}=-K^{\dagger}(\lambda).
\end{eqnarray}
In the case of the free field theory, the flow equation of the Hamiltonian operators can be rewritten as the flow equation of the coupling matrix, $\Omega^2(\lambda)$,
\begin{eqnarray}
\label{eq:the_flow_equation_omega}
\frac{d\Omega^2(\lambda)}{d\lambda}=\left[K(\lambda),\Omega^2(\lambda)\right].
\end{eqnarray}
To obtain the block diagonal Hamiltonian, we choose the generator as discussed in Eq. (\ref{eq:the_generator_choice_1}),
\begin{eqnarray}
\label{eq:the_generator_3}
K(\lambda)=\left[G(\lambda),\Omega^2(\lambda)\right],
\end{eqnarray}
where, $G(\lambda)$ is the part of the matrix $\Omega^2(\lambda)$ that the different scale coupling terms are not present. This choice ensures the anti-hermitian property of the generator as demonstrated in Eq. (\ref{eq:the_generator_2}). 

We construct the matrix $\Omega^2(\lambda)$ within the resolution truncated space $\mathcal{H}^2$ that is formed by taking the direct sum of resolution $0$ scaling and resolution $0$ and $1$ wavelet spaces. For this construction, the space is truncated to $0\leq x\leq 20$, and a constraint is imposed on the translation indices of the basis function to ensure it will vanish at the boundary. The flow equation is solved for this choice of $\Omega^2$.

\begin{figure}[hbt]
\begin{center}
\caption{The evaluation of the matrix $\Omega^2(\lambda)$ for different values of $\lambda$.}
\label{fig:scalar_field_theory_the_evaluation_of_matrix_for_different_values_of_lambda}
\includegraphics[scale=0.35]{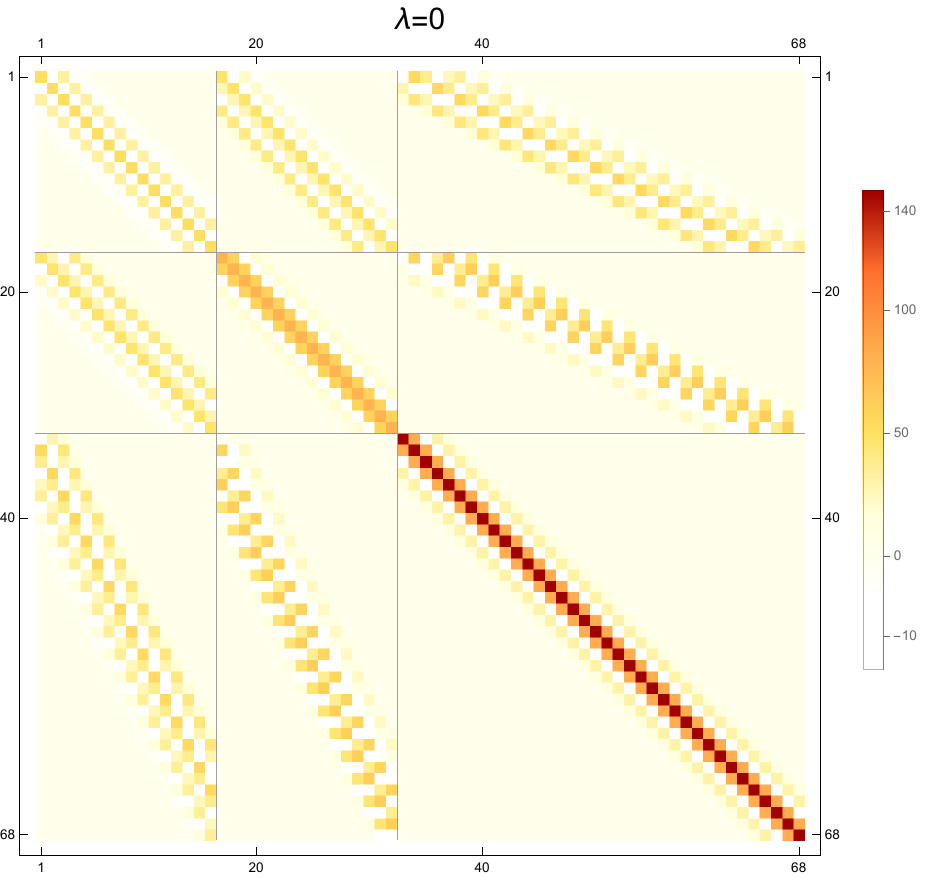}
\includegraphics[scale=0.35]{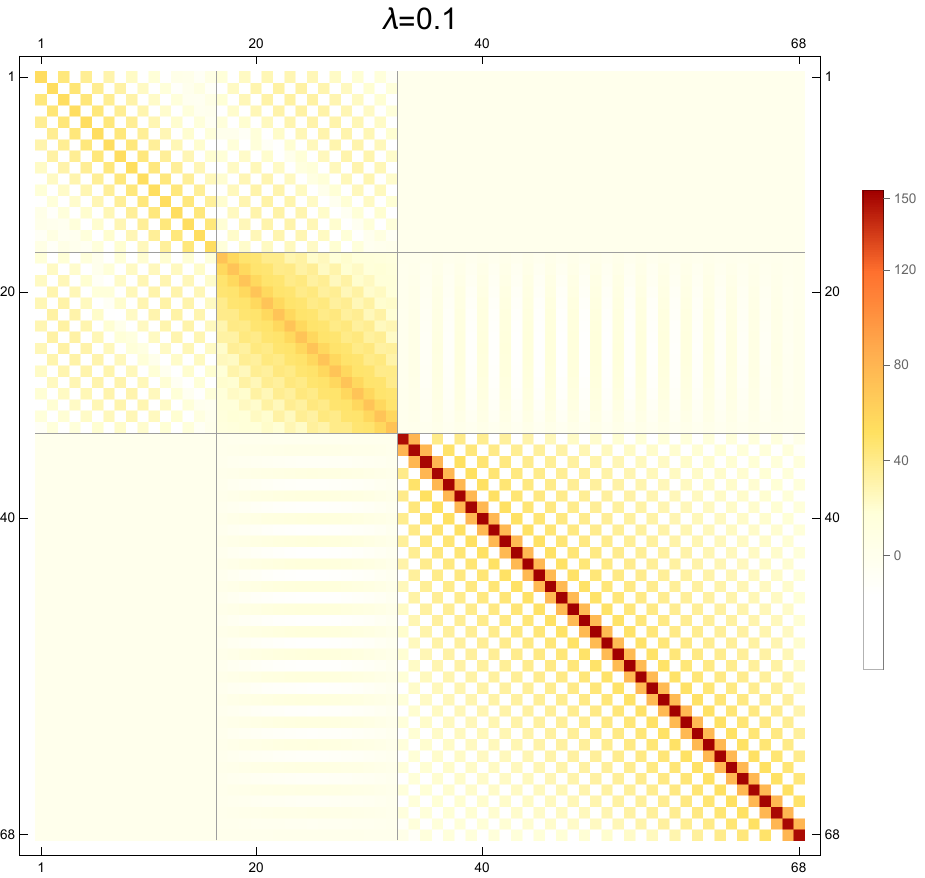}
\includegraphics[scale=0.35]{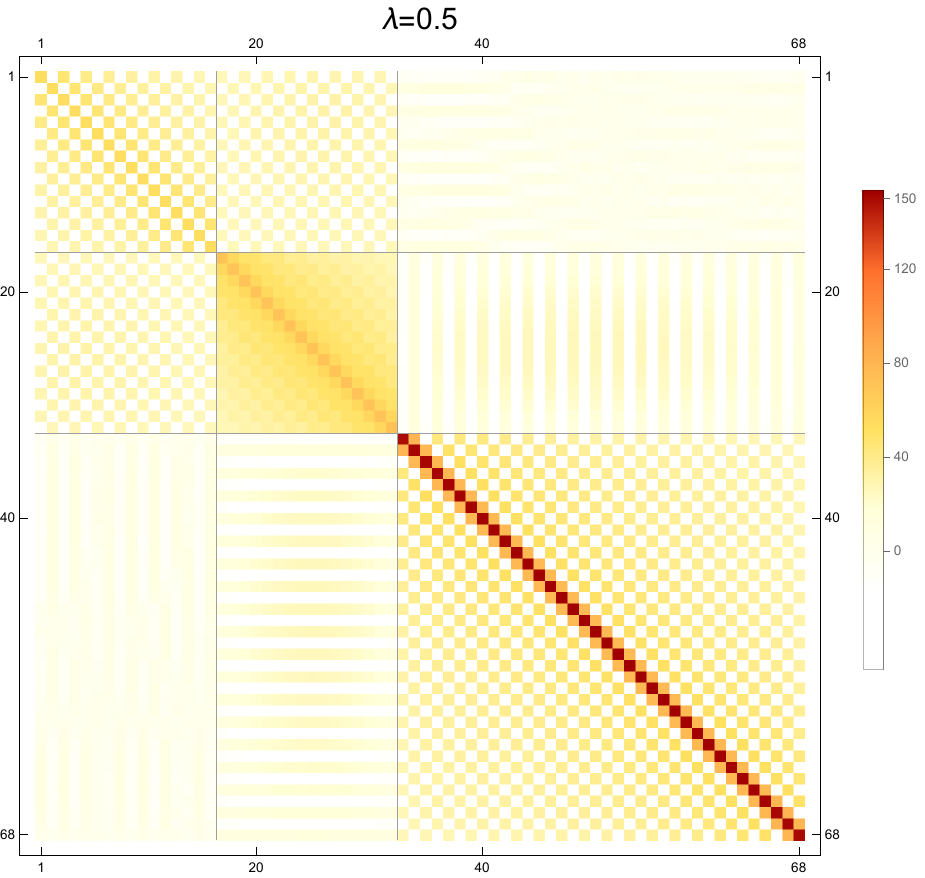}
\includegraphics[scale=0.35]{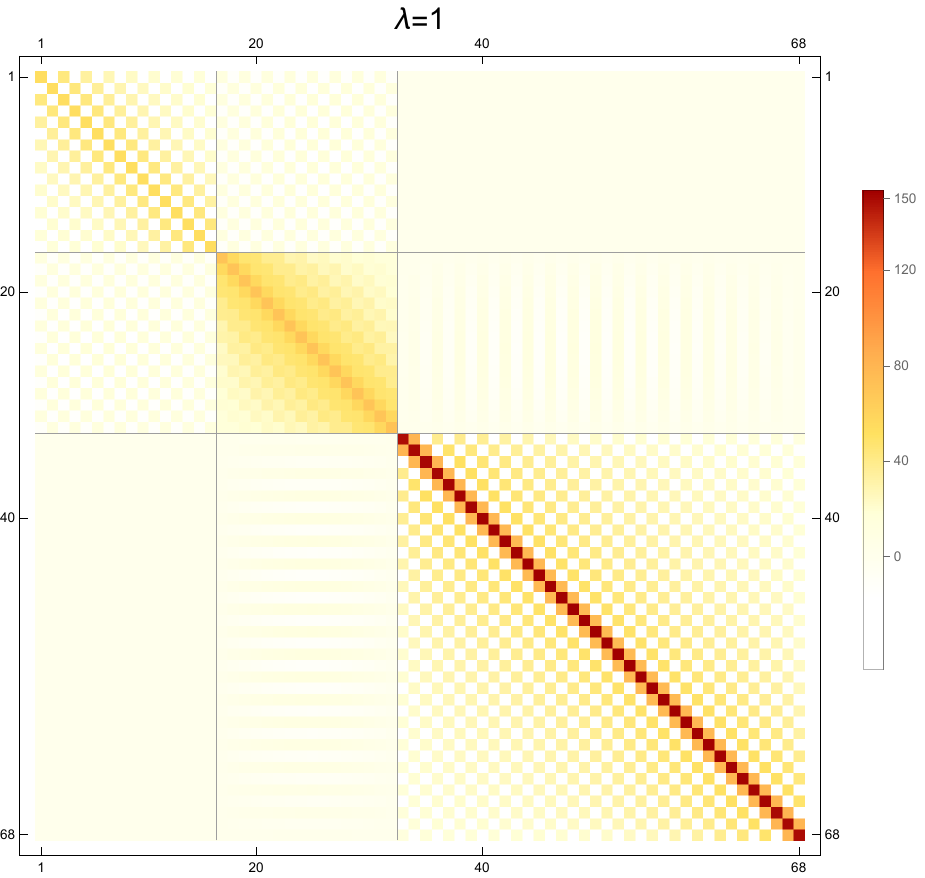}
\includegraphics[scale=0.35]{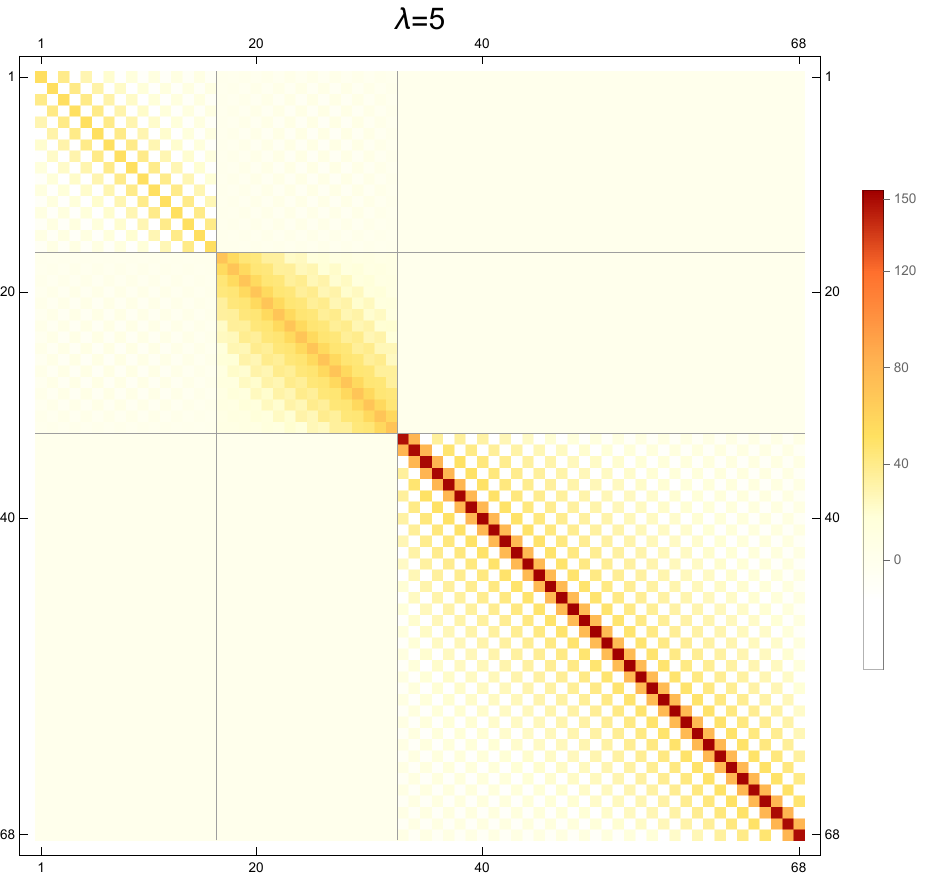}
\includegraphics[scale=0.35]{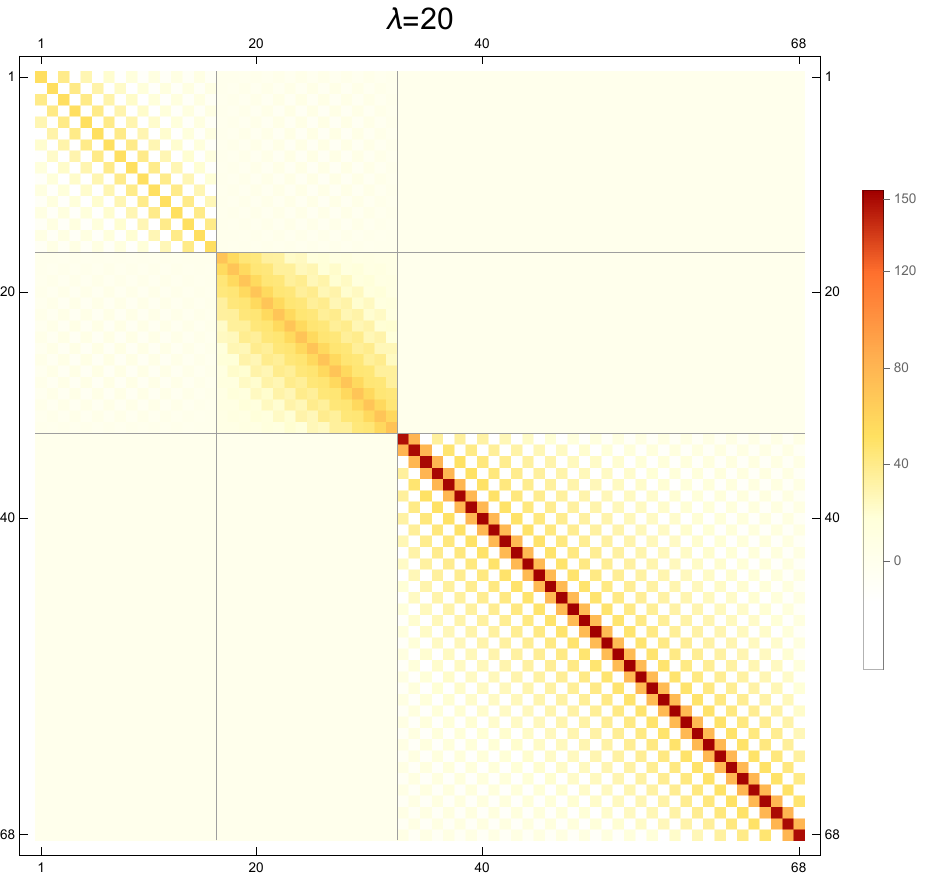}
\end{center}
\end{figure}

\begin{figure}[H]
\begin{center}
\caption{The schematic diagram of the scale coupling matrix, $\Omega^2$, truncated at resolution $2$,}
\label{fig:Schematic_Hamiltonian_k_2}
\includegraphics[scale=0.42]{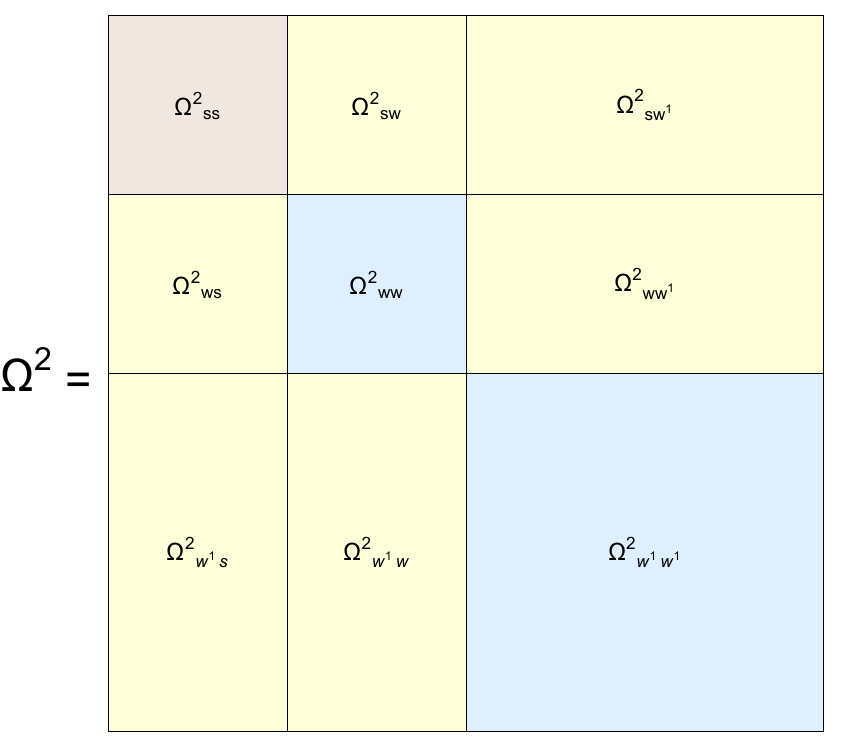}
\end{center}
\end{figure}

\begin{figure}[H]
\begin{center}
\caption{The Hilbert-Schmidt norm for six types of different quadratic expressions corresponds to six unique blocks present in $\Omega^2(\lambda)$ of free scalar field theory as a function of the flow-parameter $\lambda$.}
\label{fig:Hilbert_Schmidt_norm_for_the_scalar_field_theory}
\includegraphics[scale=0.41]{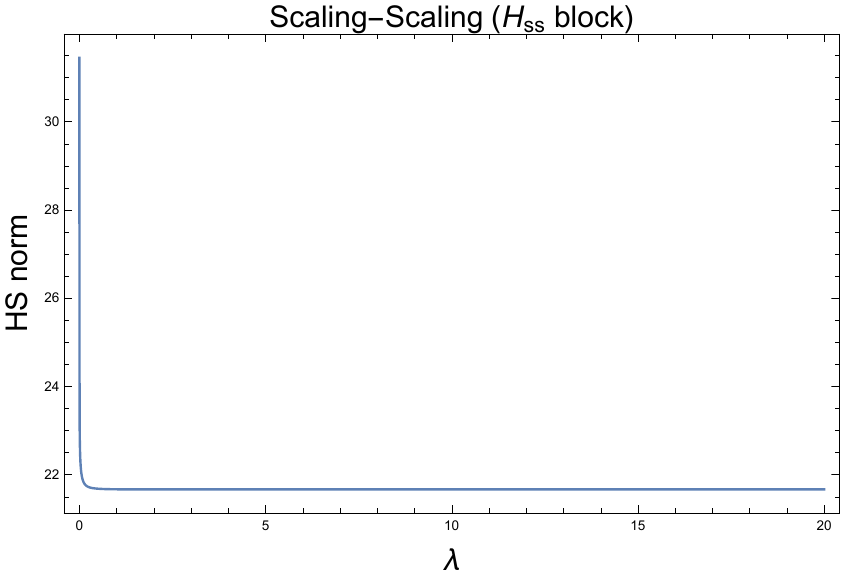}
\includegraphics[scale=0.41]{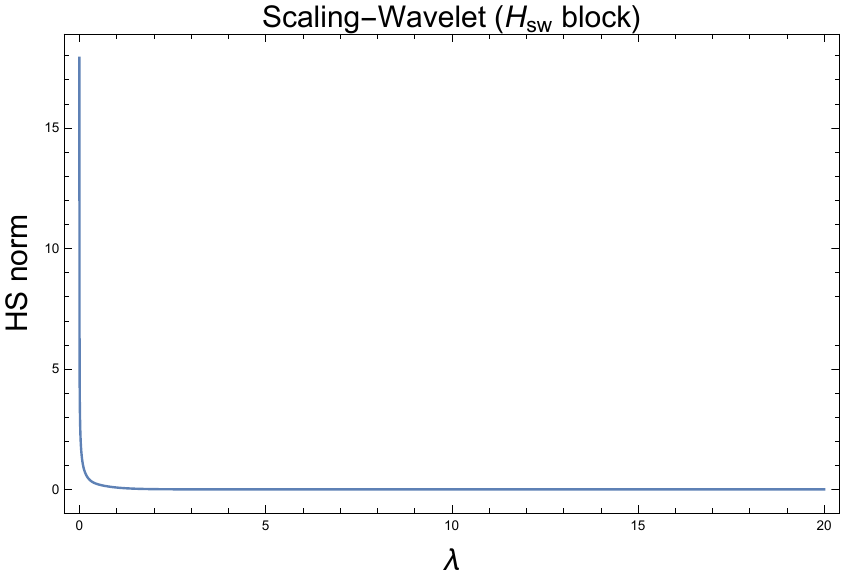}
\includegraphics[scale=0.41]{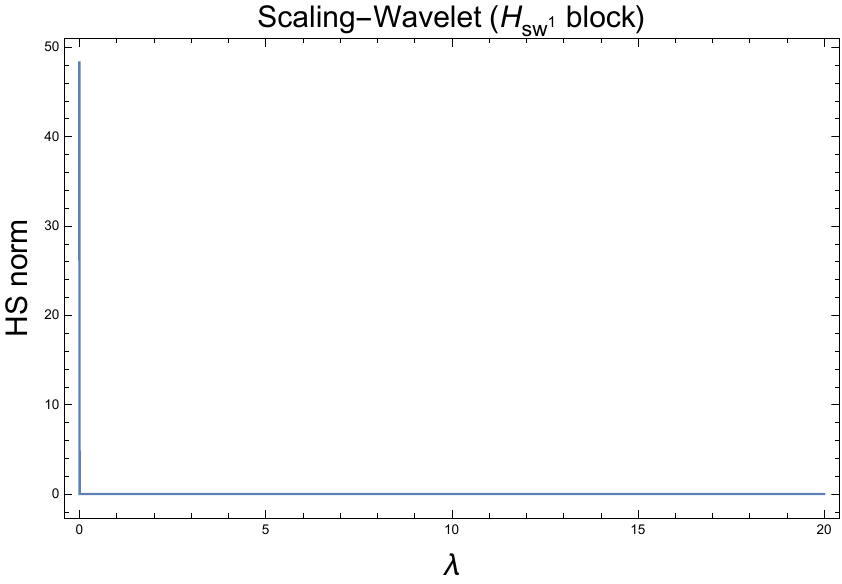}
\includegraphics[scale=0.41]{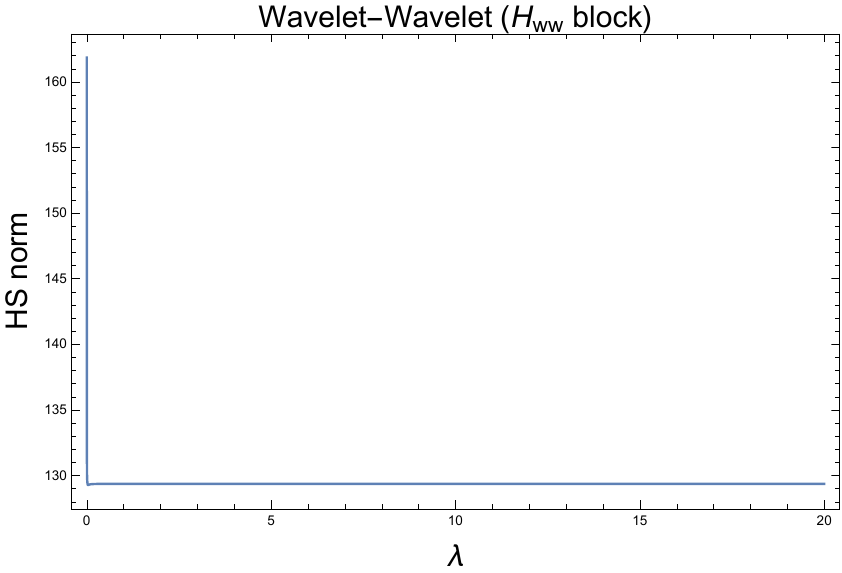}
\includegraphics[scale=0.41]{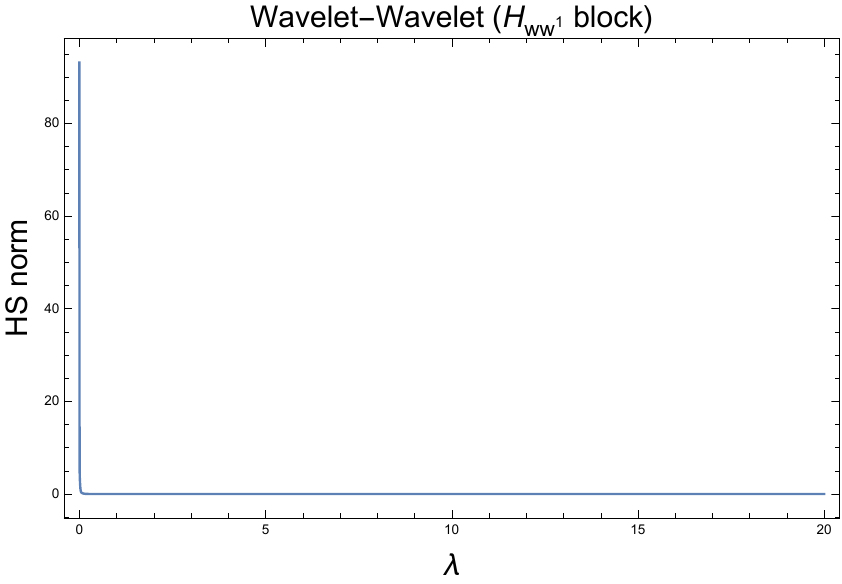}
\includegraphics[scale=0.41]{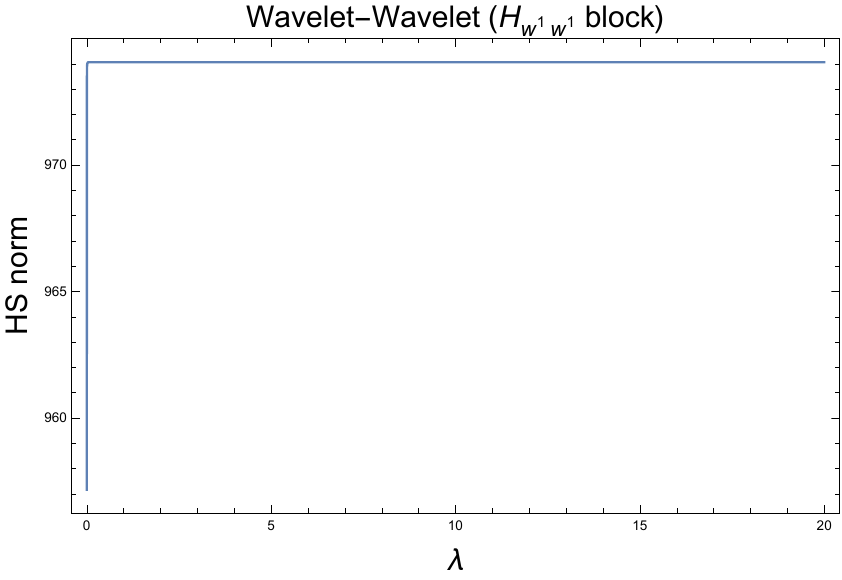}
\end{center}
\end{figure}

 We present the results of the flow equation generated $\Omega^2$ for increasing values of $\lambda$ using the matrix plots given in Fig. \ref{fig:scalar_field_theory_the_evaluation_of_matrix_for_different_values_of_lambda}. 
Each matrix plot is organised by increasing resolution as shown in Fig. The diagonal blocks, $\Omega^2_{ss}$, $\omega^2_{ww}$ and $\Omega^2_{w^1 w^1}$, represent couplings with scaling space $k=0$, wavelet space $k=0$ and wavelet space $k=1$ respectively. The off-diagonal blocks represent couplings between these resolution spaces. We observe that as the value of $\lambda$ begins to increase from $0$, the coupling between different resolutions weakens, and the matrix approaches a block diagonal form. This is validated quantitatively in Fig. Fig. \ref{fig:Hilbert_Schmidt_norm_for_the_scalar_field_theory} by computing the Hilbert-Schmidt norms (HS norm), 
\begin{eqnarray}
\label{eq:Hilbert-Schmidt_norm_1}
\sqrt{\sum_{i,j}\Omega^{2*}_{xy,ij}(\lambda)\Omega^2_{xy,ij}(\lambda)}.
\end{eqnarray}
corresponding to each block as a function of $\lambda$. In Eq. (\ref{eq:Hilbert-Schmidt_norm_1}), $\Omega^2_{xy,ij}$ represents the $(i,j)$ matrix element in the $H_{xy}$ block, where $x$ and $y$ range over the block labels (see Fig. \ref{fig:Schematic_Hamiltonian_k_2}). 
From Fig. \ref{fig:Hilbert_Schmidt_norm_for_the_scalar_field_theory}, we see that the HS norms for the off-diagonal blocks vanish with increasing $\lambda$ signifying scale decomposition. The nontrivial saturation of the HS norms for a given resolution block indicates that the effects of other resolutions have been effectively incorporated into it with increasing $\lambda$.

We further justify this observation by comparing the low-lying normal mode frequencies of the $k=2$ truncated model with those computed by diagonalising the $\Omega^2_{ss}$ block for different values of $\lambda$ in Table. \ref{tab:eigenvalues_scalar_field_theory_omega_different_values_of_lambda}.
\begin{table}[H]
\begin{center}
\caption{Comparison of the square of the normal mode frequencies of the scalar field theory calculated within the wavelet-based formalism by diagonalizing the Hamiltonian truncated at resolution $2$ and by diagonalizing the scaling function coefficient block of the Hamiltonian for different values of the flow-parameter $\lambda$.}
\label{tab:eigenvalues_scalar_field_theory_omega_different_values_of_lambda}
\setlength{\tabcolsep}{0.5pc}
\vspace{1mm}
\begin{tabular}{c | c | c | c | c | c | c}
\specialrule{.15em}{.0em}{.15em}
\hline
\multicolumn{7}{c}{Comparison of the values of $\omega_p^2$}\\
\hline
Exact & $\lambda=0$ & $\lambda=0.1$ & $\lambda=0.5$ & $\lambda=1$ & $\lambda=5$ & $\lambda=20$\\
\hline
$1.03581$ & $1.03621$ & $1.03581$ & $1.03581$ & $1.03581$ & $1.03581$ & $1.03581$\\
$1.14343$ & $1.14567$ & $1.14343$ & $1.14343$ & $1.14343$ & $1.14343$ & $1.14343$\\
$1.32351$ & $1.32351$ & $1.32351$ & $1.32351$ & $1.32351$ & $1.32351$ & $1.32351$\\
$1.57716$ & $1.57716$ & $1.57716$ & $1.57716$ & $1.57716$ & $1.57716$ & $1.57716$\\
$1.90561$ & $1.90561$ & $1.90561$ & $1.90561$ & $1.90561$ & $1.90561$ & $1.90561$\\
$2.30971$ & $2.30971$ & $2.30971$ & $2.30971$ & $2.30971$ & $2.30971$ & $2.30971$\\
$2.78938$ & $2.78938$ & $2.78938$ & $2.78938$ & $2.78938$ & $2.78938$ & $2.78938$\\
$3.34334$ & $3.34334$ & $3.34334$ & $3.34334$ & $3.34334$ & $3.34334$ & $3.34334$\\
$3.96959$ & $3.96959$ & $3.96959$ & $3.96959$ & $3.96959$ & $3.96959$ & $3.96959$\\
$4.66730$ & $4.66730$ & $4.66730$ & $4.66730$ & $4.66730$ & $4.66730$ & $4.66730$\\
$5.43990$ & $5.43990$ & $5.43990$ & $5.43990$ & $5.43990$ & $5.43990$ & $5.43990$\\
$6.29676$ & $6.29677$ & $6.29677$ & $6.29676$ & $6.29676$ & $6.29676$ & $6.29676$\\
$7.25035$ & $7.25054$ & $7.25054$ & $7.25035$ & $7.25035$ & $7.25035$ & $7.25035$\\
$8.31159$ & $8.32024$ & $8.32024$ & $8.31159$ & $8.31159$ & $8.31159$ & $8.31159$\\
$9.48841$ & $9.72486$ & $9.72486$ & $9.48936$ & $9.48841$ & $9.48841$ & $9.48841$\\
$10.7869$ & $10.9552$ & $10.9552$ & $10.8170$ & $10.7910$ & $10.7869$ & $10.7869$\\
\hline
\specialrule{.15em}{.15em}{.0em}
\end{tabular}
\end{center}
\end{table}
The accuracy of the square of the normal mode frequencies, calculated by diagonalizing the $\Omega^2_{ss}$ block of the equivalent Hamiltonian with physically relevant degrees of freedom (scaling function degrees of freedom), reduces from $10^{-2}$ to $10^{-9}$ as the flow , $\lambda$, increases from $0$ to $20$.

This analysis demonstrates that the flow-equation method evolves the initial truncated Hamiltonian to a block diagonal form, with blocks organized by resolution. This form of the Hamiltonian may be particularly useful in scenarios where multiple resolutions contribute to the physics of the model.

\section{Quantization of the model of two interacting scalar fields: Fourier basis}
\label{sec:model_of_two_interacting_scalar_fields}
In this section, we describe a model of two scalar fields, $\phi$ and $\psi$, with masses 
$\mu$ and $\nu$, in $1+1$ dimensions, interacting via a mass mixing coupling $g$. 
This model will be used to study the separation of low- and high-resolution degrees of freedom in the Hamiltonian. The Lagrangian density of this chosen model is given by,
\begin{eqnarray}
\label{eq:model_lagrangian_density}
&&\mathcal{L}=\frac{1}{2}\left[\dot{\phi}^2(x,t)-(\partial_x \phi(x,t))^2-\mu^2 \phi(x,t)^2\right]+\frac{1}{2}\left[\dot{\psi}^2(x,t)-(\partial_x \psi(x,t))^2-\nu^2\psi(x,t)^2 \right] -\frac{g^2}{2} \phi(x,t) \psi(x,t).
\end{eqnarray}
The corresponding Hamiltonian has the form, 
\begin{eqnarray}
\mathrm{H}=\int dx\left(\frac{1}{2}\left[\pi_{\phi}^2+\left(\partial_x \phi \right)^2+\mu^2\phi^2\right]+\frac{1}{2}\left[\pi_{\psi}^2+\left(\partial \psi\right)^2+\nu^2\psi^2\right]+\frac{g^2}{2}\phi\psi\right).\quad
\end{eqnarray}
where,
\begin{eqnarray}
\pi_\phi=\frac{\partial \mathcal{L}}{\partial \dot{\phi}}=\dot{\phi}, \,\text{and}\quad \pi_\psi=\frac{\partial \mathcal{L}}{\partial \dot{\psi}}=\dot{\psi}
\end{eqnarray}
represent the canonical momenta corresponding to $\phi$ and $\psi$ respectively.
Expanding the fields in terms of the Fourier modes within the interval $L$, as described in the previous section, transforms the Hamiltonian to,
\begin{eqnarray}
\mathrm{H}=\sum_{p=1}^{\infty}\left(\frac{1}{2}\left[\pi_{\phi,p}(t)^2+(p^2+\mu^2)\phi_p(t)^2+\pi_{\psi,p}(t)^2+(p^2+\nu^2)\psi_p(t)^2\right]+\frac{g^2}{2} \phi_p(t)\psi_p(t)\right),
\end{eqnarray}
where,
\begin{eqnarray}
\phi(\textrm{x},t)&=&\sum_{p=1}^{\infty} \phi_{p}(t)\sqrt{\frac{2}{L}}\sin\left(\frac{p\pi x}{L}\right), \quad \pi_\phi(\textrm{x},t)=\sum_{p=1}^{\infty} \pi_{\phi,p}(t)\sqrt{\frac{2}{L}}\sin\left(\frac{p\pi x}{L}\right),\\
\psi(\textrm{x},t)&=&\sum_{p=1}^{\infty} \psi_{p}(t)\sqrt{\frac{2}{L}}\sin\left(\frac{p\pi x}{L}\right), \quad \pi_\psi(\textrm{x},t)=\sum_{p=1}^{\infty} \pi_{\psi,p}(t)\sqrt{\frac{2}{L}}\sin\left(\frac{p\pi x}{L}\right).
\end{eqnarray}
The full Hamiltonian, rewritten in the momentum modes,
\begin{eqnarray}
\textrm{H}=\sum_{p=1}^{\infty}\frac{1}{2}\left[\begin{pmatrix}
\pi_{\phi,p}(t)&\pi_{\psi,p}(t)
\end{pmatrix}
\begin{pmatrix}
\mathds{1}&0\\
0&\mathds{1}
\end{pmatrix}
\begin{pmatrix}
\pi_{\phi,p}(t)\\
\pi_{\psi,p}(t)
\end{pmatrix}+ 
\begin{pmatrix}
\phi_p(t)&\psi_p(t)
\end{pmatrix}
\begin{pmatrix}
\mu^2+p^2 & \frac{g^2}{2}\\
\frac{g^2}{2}& \nu^2+p^2
\end{pmatrix}
\begin{pmatrix}
\phi_p(t)\\
\psi_p(t)
\end{pmatrix}\right],
\end{eqnarray}
shows that the mass mixing term only couples the same momentum mode of $\phi_p$ and $\psi_p$. The quadratic nature of the mass mixing term allows one to introduce the normal mode variables, 
$\Phi_p$ and $\Psi_p$, 
\begin{eqnarray}
\Phi_p(t)&=&\frac{1}{2}\left[\left(\frac{(\nu^2-g^2)-\sqrt{(\nu^2-\mu^2)^2+g^4}}{\sqrt{(\nu^2-\mu^2)^2+g^4}}\right)^{\frac{1}{2}}\phi_p(t)+\left(\frac{\sqrt{(\nu^2-\mu^2)^2+g^4}-(\nu^2-g^2)}{\sqrt{(\nu^2-\mu^2)^2+g^4}}\right)^{\frac{1}{2}}\psi_p(t)\right],\\
\Psi_p(t)&=&\frac{1}{2}\left[\left(\frac{(\nu^2-g^2)-\sqrt{(\nu^2-\mu^2)^2+g^4}}{\sqrt{(\nu^2-\mu^2)^2+g^4}}\right)^{\frac{1}{2}}\phi_p(t)-\left(\frac{\sqrt{(\nu^2-\mu^2)^2+g^4}-(\nu^2-g^2)}{\sqrt{(\nu^2-\mu^2)^2+g^4}}\right)^{\frac{1}{2}}\psi_p(t)\right],
\end{eqnarray}
in terms of which the Hamiltonian takes the form,
\begin{eqnarray}
\textrm{H}= \sum_{p=1}^{\infty}\frac{1}{2}\left[\Pi^2_{\Phi,p}(t)+\omega^2_{\Phi,p}\Phi^2_p(t)+\Pi^2_{\Psi,p}(t)+\omega^2_{\Psi,p}\Psi^2_p(t)\right]
\end{eqnarray}
where, 
\begin{eqnarray}
\label{eq:exact_omega_phi_p}
\omega_{\Phi,p}=\sqrt{\left(\frac{p\pi}{L}\right)^2+\frac{\mu^2+\nu^2}{2}+\frac{1}{2}\sqrt{(\mu^2-\nu^2)^2+g^4} }\\
\label{eq:exact_omega_psi_p}
\omega_{\Psi,p}=\sqrt{\left(\frac{p\pi}{L}\right)^2+\frac{\mu^2+\nu^2}{2}-\frac{1}{2}\sqrt{(\mu^2-\nu^2)^2+g^4}}
\end{eqnarray}
denote the corresponding normal mode frequencies. The exact expression for the normal frequencies will be used when validating the flow-equation generated effective Hamiltonian. The exact values of these normal mode frequencies, as computed from Eq. (\ref{eq:exact_omega_phi_p}) and Eq. (\ref{eq:exact_omega_psi_p}) for $\mu = \nu \ = g = 1$, are given in the first column of Table. \ref{tab:compare_omega_and_omega_for_resolution_10}.

\section{Quantization of the model of two interacting scalar fields: Wavelet basis}
\label{sec:Canonical_Quantization_in_wavelet_basis}
In this section, we outline the wavelet-based canonical quantization of the model defined in Sec. \ref{sec:model_of_two_interacting_scalar_fields}. Since the approach is similar to the one followed in Sec. \ref{sec:The_scalar_field_theory_and_the_normal_mode_frequencies} \cite{PhysRevD.95.094501}, we only provide the essential features of the procedure. We begin by resolving both the scalar fields, $\phi$ and $\psi$, and their canonical conjugates, $\pi_\phi$ and $\pi_\psi$, in Daubechies wavelet basis,
\begin{eqnarray}
\label{eq:discretized_phi_field}
\phi(x,t)&=&\sum_{n}\phi^{s,k}_{n}(t)s^{k}_n(x)+\sum_{n,l\ge k}\phi^{w,l}_{n}(t)w^l_n(x),\\
\label{eq:discretized_pi_phi_field}
\pi_{\phi}(x,t)&=&\sum_{n}\pi^{s,k}_{\phi ,n}(t)s^k_n(x)+\sum_{n,l\ge k}\pi^{w,l}_{\phi, n}(t)w^l_n(x),
\end{eqnarray}
and,
\begin{eqnarray}
\label{eq:discretized_psi_field}
\psi(x,t)&=&\sum_{n}\psi^{s,k}_{n}(t)s^k_n(x)+\sum_{n,l\ge k}\psi^{w,l}_{n}(t)w^l_n(x),\\
\label{eq:discretized_pi_psi_field}
\pi_{\psi}(x,t)&=&\sum_{n}\pi^{s,k}_{\psi, n}(t)s^k_n(x)+\sum_{n,l\ge k}\pi^{w,l}_{\psi, n}(t)w^l_n(x).
\end{eqnarray}
The discrete scaling and wavelet basis coefficients can be extracted from the fields and their canonical conjugates using,
\begin{eqnarray}
\begin{aligned}
\label{eq:definition_of_phi_coefficients}
\phi^{s,k}_{n}(t)&=&\int dx \phi(x,t)s^k_n(x),\;\;\;
\phi^{w,l}_{n}(t)&=&\int dx \phi(x,t)w^l_n(x),\\
\pi^{s,k}_{\phi, n}(t)&=&\int dx \pi_\phi(x,t)s^k_n(x),\;\;\;
\pi^{w,k}_{\phi, n}(t)&=&\int dx \pi_\phi(x,t)w^l_n(x),
\end{aligned}
\end{eqnarray}
and
\begin{eqnarray}
\begin{aligned}
\label{eq:definition_of_psi_coefficients}
\psi^{s,k}_{n}(t)&=&\int dx \psi(x,t)s^k_n(x),\;\;\;
\psi^{w,l}_{n}(t)&=&\int dx \psi(x,t)w^l_n(x),\\
\pi^{s,k}_{\psi n}(t)&=&\int dx \pi_\psi(x,t)s^k_n(x),\;\;\;
\pi^{w,l}_{\psi, n}(t)&=&\int dx \pi_\psi(x,t)w^l_n(x).
\end{aligned}
\end{eqnarray}
The equal time canonical commutation relation of the fields and their conjugates,
\begin{eqnarray}
\begin{aligned}
\left[\phi(x,t),\pi_\phi(y,t)\right]=i \delta(x-y),&&\quad
\left[\phi(x,t),\phi(y,t)\right]=0,\\
\left[\psi(x,t),\pi_\psi(y,t)\right]=i \delta(x-y),&&\quad
\left[\psi(x,t),\psi(y,t)\right]=0,
\end{aligned}
\end{eqnarray}
written in terms of the scaling and wavelet basis coefficients, using Eq. (\ref{eq:definition_of_phi_coefficients}) and Eq. (\ref{eq:definition_of_psi_coefficients}), reveal as mutually independent canonical pairs of phase space variables. The nontrivial commutation relations are given by,
\begin{widetext}
\begin{eqnarray}
\begin{aligned}
\left[\phi^{s,k}_n(t),\pi^{s,k}_{\phi,m}(t)\right]=i \delta_{mn},\quad
\left[\phi^{w,l}_{n}(t),\pi^{w,r}_{\phi,m}(t)\right]=i \delta_{mn}\delta_{lr}\\
\end{aligned}
\end{eqnarray}
and
\begin{eqnarray}
\begin{aligned}
\left[\psi^{s,k}_n(t),\pi^{s,k}_{\psi,m}(t)\right]=i \delta_{mn},\quad
\left[\psi^{w,l}_{n}(t),\pi^{w,r}_{\psi,n}(t)\right]=i \delta_{mn}\delta_{lr},\\
\end{aligned}
\end{eqnarray}
\end{widetext}
while the rest of the pair combinations commute with each other. 

We express the Hamiltonian in terms of the scaling and wavelet canonical variables by substituting Eq. (\ref{eq:discretized_phi_field})-Eq. (\ref{eq:discretized_pi_psi_field}) into Eq. (\ref{eq:model_lagrangian_density}). The result shows that the Hamiltonian is a sum of three terms, 
\begin{eqnarray}
\label{eq:phipsi_hamiltonian}
\mathrm{H}=\mathrm{H}_{ss}+\mathrm{H}_{ww}+\mathrm{H}_{sw}
\end{eqnarray}
where $\mathrm{H}_{ss}$ describes the physics of the model on coarse length scales down to resolution $k$,
\begin{eqnarray}
\label{eq:phipsi_h_ss}
\mathrm{H}_{ss}:=&&\frac{1}{2}\left(\sum_n \pi^{s,k}_{\phi,n}\pi^{s,k}_{\phi,n}+\sum_n \pi^{s,k}_{\psi,n}\pi^{s,k}_{\psi,n}
+\sum_{n,m}\phi^{s,k}_{n}\phi^{s,k}_{m}\mathcal{D}^k_{ss,nm}+\sum_{n,m}\psi^{s,k}_{n}\psi^{s,k}_{m}\mathcal{D}^k_{s,nm}\right.\nonumber\\
&&\left.+\sum_n \mu^2 \phi^{s,k}_{n}\phi^{s,k}_{n}+\sum_n \nu^2 \psi^{s,k}_{n}\psi^{s,k}_{n}
+\sum_n g^2 \phi^{s,k}_{n}\psi^{s,k}_{n}\right),
\end{eqnarray}
$\mathrm{H}_{ww}$ describes the physics of the model on length scales finer than resolution $k$, 
\begin{eqnarray}
\label{eq:phipsi_h_ww}
\mathrm{H}_{ww}:=&&\frac{1}{2}\left(\sum_{\substack{n \\ l\ge k}} \pi^{w,l}_{\phi,n}\pi^{w,l}_{\phi,n}+\sum_{\substack{n \\l\ge k}} \pi^{w,l}_{\psi,n}\pi^{w,l}_{\psi,n}
+\sum_{\substack{n,m \\(l,q)\ge k}}\phi^{w,l}_{n}\phi^{w,q}_{m}\mathcal{D}^{lq}_{ww,mn}+\sum_{\substack{n,m \\(l,q)\ge k}}\psi^{w,l}_{n}\psi^{w,q}_{m}\mathcal{D}^{lq}_{ww,mn}\right.\nonumber\\
&&\left.+\sum_{\substack{n \\ {l}\ge k}} \mu^2 \phi^{w,l}_{n}\phi^{w,l}_{n}+\sum_{\substack{n \\ l\ge k}} \nu^2 \psi^{w,l}_{n}\psi^{w,l}_{n}
+\sum_{\substack{n \\ l\ge k}} g^2 \phi^{w,l}_{n}\psi^{w,l}_{n}\right)
\end{eqnarray}
and, $\mathrm{H}_{sw}$ represents interactions between lengths scales coarser and finer than resolution $k$,
\begin{eqnarray}
\label{eq:phipsi_h_sw}
\mathrm{H}_{sw}:=&&\frac{1}{2}\left(\sum_{\substack{n,m \\ q\ge k}} \phi^k_{n}\phi^{w,q}_{m}\mathcal{D}^{kq}_{sw,nm}+\sum_{\substack{n,m \\ l\ge k}} \phi^{w,l}_{m}\phi^{s,k}_{n}\mathcal{D}^{kl}_{sw,mn} 
+\sum_{\substack{n,m \\ q\ge k}} \psi^{s,k}_{n}\psi^{w,q}_{m}\mathcal{D}^{kq}_{sw,nm}+\sum_{\substack{n,m \\ l\ge k}} \psi^{w,l}_{m}\psi^k_{n}\mathcal{D}^{kl}_{sw,mn}\right).
\end{eqnarray}
The form of $\mathrm{H}_{sw}$ shows that the nature of the coupling between $\phi$ and $\psi$ is such that they don't couple across different length scales.  The coefficients $\mathcal{D}^k_{ss,nm}$, $\mathcal{D}^{lq}_{ww,nm}$ and $\mathcal{D}^{kq}_{sw,nm}$ are the constant matrices given by,
\begin{eqnarray}
\mathcal{D}^k_{ss,nm}&=&\int \frac{d}{dx}s^k_n(x)\frac{d}{dx}s^k_m(x)dx\\
\mathcal{D}^{lq}_{ww,nm}&=&\int \frac{d}{dx}w^l_n(x)\frac{d}{dx}w^q_m(x)dx\\
\mathcal{D}^{kq}_{sw,nm}&=&\int \frac{d}{dx}s^k_n(x)\frac{d}{dx}w^q_m(x)dx \;\;\; q\ge k 
\end{eqnarray}
These matrices can be evaluated using the procedure due to Beylkin \cite{10.2307/2158264}. This procedure is outlined in detail by Polyzou \cite{PhysRevD.87.116011}. 

\section{Analysis}
\label{sec:analysis}
 Using the model defined in Sec \ref{sec:model_of_two_interacting_scalar_fields}, we demonstrate the decoupling of the low resolution from the high resolution degrees of freedom in a theory with a generally quadratic interaction using the flow equation method. We also show that the low energy physics predicted by the effective Hamiltonian contains effects of both low and high-resolution variables. This constitutes an extension of the analysis of the free scalar field theory due to Michlin and Polyzou \cite{PhysRevD.95.094501}. 


The Hamiltonian of this quantum field theory, Eq. (\ref{eq:phipsi_h_ss})-Eq. (\ref{eq:phipsi_h_ww}), can be rewritten in a compact form,
\begin{eqnarray}
&&\mathrm{H}=\sum_{m,n=-\infty}^{\infty}\frac{1}{2}\left[\begin{pmatrix}
\pi^{s,k}_{\phi,m}\;&\pi^{s,k}_{\psi,m}\;&\pi^{w,l}_{\phi,m}\;&\pi^{w,l}_{\psi,m}
\end{pmatrix}
\begin{pmatrix}
\delta_{mn}&0&0&0\\
0&\delta_{mn}&0&0\\
0&0&\delta_{mn}&0\\
0&0&0&\delta_{mn}\\
\end{pmatrix}
\begin{pmatrix}
\pi^{s,k}_{\phi,n}\\
\pi^{s,k}_{\psi,n}\\
\pi^{w,q}_{\phi,n}\\
\pi^{w,q}_{\psi,n}
\end{pmatrix}+ \right.\nonumber \\
\label{eq:psiphi_hamiltonian_matrix_form}
&&\left.\begin{pmatrix}
\phi^{s,k}_{m} & \psi^{s,k}_{m} & \phi^{w,l}_{m} & \psi^{w,l}_{m}
\end{pmatrix}
\underbrace{\left(
\begin{array}{@{}c|c@{}}
\begin{matrix}
\mu^2\delta_{mn}+\mathcal{D}^k_{ss,mn} & \frac{g^2}{2}\delta_{mn} \\
\frac{g^2}{2}\delta_{mn} & \nu^2\delta_{mn}+\mathcal{D}^k_{ss,mn}
\end{matrix}
&
\begin{matrix}
\mathcal{D}^{kq}_{sw,mn} & 0 \\
0 & \mathcal{D}^{kq}_{sw,mn}
\end{matrix}\\
\cmidrule[0.4pt]{1-2}\\
\begin{matrix}
\mathcal{D}^{qk}_{ws,mn} & 0 \\
0 & \mathcal{D}^{qk}_{ws,mn}
\end{matrix}
&
\begin{matrix}
\mu^2\delta_{mn}+\mathcal{D}^{lq}_{ww,mn} & \frac{g^2}{2}\delta_{mn}\\
\frac{g^2}{2}\delta_{mn} & \nu^2\delta_{mn}+\mathcal{D}^{lq}_{ww,mn}
\end{matrix}
\end{array}
\right)}_{\Omega^2}
\begin{pmatrix}
\phi^{s,k}_{n} \\ \psi^{s,k}_{n} \\ \phi^{w,q}_{n} \\ \psi^{w,q}_{n}
\end{pmatrix}
\right]\quad
\end{eqnarray}
where, $\Omega^2$ denotes a matrix that represents the coupling of each scalar field across length scales. From Eq. (\ref{eq:psiphi_hamiltonian_matrix_form}), we see that the two scalar fields couple to each other only on the same scale. They do not have any coupling across scales.

The symmetric nature of the coupling matrix $\Omega^2$ implies the existence of an orthogonal matrix $O$, using which the normal mode frequencies and the corresponding normal mode variables can be computed. There exists a unitary operator $U$ that generates this orthogonal transformation $O$,
\begin{eqnarray}
\label{eq:the_transformation_of_filed_operators}
U
\begin{pmatrix}
\phi^{s,k} \\
\psi^{s,k} \\
\phi^{w,l} \\
\psi^{w,l}
\end{pmatrix}
U^{\dagger}
=
\begin{pmatrix}
\Phi^{s,k} \\
\Psi^{s,k} \\
\Phi^{w,l} \\
\Psi^{w,l}
\end{pmatrix}
=O
\begin{pmatrix}
\phi^{s,k} \\
\psi^{s,k} \\
\phi^{w,l} \\
\psi^{w,l}
\end{pmatrix}
\quad \text{and}\quad
U
\begin{pmatrix}
\pi^{s,k} \\
\pi^{s,k} \\
\pi^{w,l} \\
\pi^{w,l}
\end{pmatrix}
U^{\dagger}
=
\begin{pmatrix}
\Pi^{s,k} \\
\Pi^{s,k} \\
\Pi^{w,l} \\
\Pi^{w,l}
\end{pmatrix}
=O
\begin{pmatrix}
\pi^{s,k} \\
\pi^{s,k} \\
\pi^{w,l} \\
\pi^{w,l}
\end{pmatrix}.
\end{eqnarray}
Here, the superscripts on the normal mode variables $\Phi^{s,k}$, $\Phi^{w,l}$, $\Psi^{s,k}$ and $\Psi^{w,l}$, serve as a mere reminder that they arose from $\phi^{s,k}$, $\phi^{w,l}$, $\psi^{s,k}$ and $\psi^{w,l}$ through an orthogonal rotation. The rotation $O$ itself mixes resolutions. The unitary transform of the Hamiltonian gives the normal mode representation of the model,
\begin{eqnarray}
\mathrm{H}'=U\mathrm{H}U^{\dagger}=&&\frac{1}{2}
\left[
\begin{pmatrix}
\Pi^{s,k} & \Pi^{s,k} & \Pi^{w,l} & \Pi^{w,l}
\end{pmatrix}
\begin{pmatrix}
\mathds{1}&0&0&0\\
0&\mathds{1}&0&0\\
0&0&\mathds{1}&0\\
0&0&0&\mathds{1}
\end{pmatrix}
\begin{pmatrix}
\Pi^{s,k} \\
\Pi^{s,k} \\
\Pi^{w,l} \\
\Pi^{w,l}
\end{pmatrix}\right.\nonumber\\
&&+
\left.\begin{pmatrix}
\Phi^{s,k} & \Psi^{s,k} & \Phi^{w,l} & \Psi^{w,l}
\end{pmatrix}
\begin{pmatrix}
\omega^2_{s,\Phi} & 0 & 0 & 0\\
0 & \omega^2_{s,\Psi} & 0 & 0\\
0 & 0 & \omega^2_{w,\Phi} & 0\\
0 & 0 & 0 & \omega^2_{w,\Psi}
\end{pmatrix}
\begin{pmatrix}
\Phi^{s,k} \\
\Psi^{s,k} \\
\Phi^{w,l} \\
\Psi^{w,l}
\end{pmatrix}
\right],
\end{eqnarray}
where, $\omega^2_{s,\Phi}$, $\omega^2_{s,\Phi}$, $\omega^2_{w,\Psi}$, and $\omega^2_{w,\Psi}$ are the diagonal matrices containing the squares of the normal mode frequencies associated with $\Phi$ and $\Psi$ respectively. The correspondence between the normal mode variables and normal mode frequencies in the Fourier based analysis given in Sec. \ref{sec:model_of_two_interacting_scalar_fields} and the present wavelet based analysis is given by,
\begin{eqnarray}
\begin{pmatrix}
\Phi^{s,k} \\
\Phi^{w,l}
\end{pmatrix}
\rightarrow
\Phi_p,
\quad
\begin{pmatrix}
\Pi^{s,k}_{\Phi} \\
\Pi^{w,l}_{\Phi}
\end{pmatrix}
\rightarrow
\Pi_{\Phi,p},
\quad
\begin{pmatrix}
\omega^2_{s,\Phi} \\
\omega^2_{w,\Phi}
\end{pmatrix}
\rightarrow
\omega^2_{\Phi,p}\\
\begin{pmatrix}
\Psi^{s,k} \\
\Psi^{w,l}
\end{pmatrix}
\rightarrow
\Psi_p,
\quad
\begin{pmatrix}
\Pi^{s,k}_{\Psi} \\
\Pi^{w,l}_{\Psi}
\end{pmatrix}
\rightarrow
\Pi_{\Psi,p},
\quad
\begin{pmatrix}
\omega^2_{s,\Psi} \\
\omega^2_{w,\Psi}
\end{pmatrix}
\rightarrow
\omega^2_{\Psi,p}
\end{eqnarray}

As an illustration, in Table \ref{tab:compare_omega_and_omega_for_resolution_10}, we present a comparison between the exact model and the model truncated to resolutions $k=1$, $k=6$ and $k=10$ for the first sixteen normal mode frequencies. The parameter $\mu$, $\nu$ and $g$ are set to $1$. As in the case of the free theory of a single real scalar field, we use pure scaling function basis representation to calculate the normal mode frequencies.
\begin{table}[H]
\begin{center}
\caption{Comparison between the exact and wavelet-based estimation of the square of the normal mode frequencies for different resolutions }
\label{tab:compare_omega_and_omega_for_resolution_10}
\setlength{\tabcolsep}{0.5pc}
\vspace{1mm}
\begin{tabular}{c  c | c  c | c  c | c c | c c }
\specialrule{.15em}{.0em}{.15em}
\hline
\multicolumn{2}{c|}{Exact $\omega_p^2$} & \multicolumn{2}{c|}{$\omega_p^2$ for $k=0$} & \multicolumn{2}{c|}{$\omega_p^2$ for $k=1$} & \multicolumn{2}{c|}{$\omega^2_p$ for $k=6$} & \multicolumn{2}{c}{$\omega^2_p$ for $k=10$} \\
\hline
$\Phi$ & $\Psi$ & $\Phi$ & $\Psi$ & $\Phi$ & $\Psi$ & $\Phi$ & $\Psi$ & $\Phi$ & $\Psi$ \\
\hline
$0.52467$ & $1.52467$ & $0.53621$ & $1.53621$ & $0.52960$ & $1.52960$ & $0.52481$ & $1.52481$ & $0.52468$ & $1.52468$\\
$0.59870$ & $1.59870$ & $0.64567$ & $1.64567$ & $0.61846$ & $1.61846$ & $0.59924$ & $1.59924$ & $0.59873$ & $1.59873$\\
$0.72207$ & $1.72207$ & $0.83255$ & $1.83255$ & $0.76681$ & $1.76681$ & $0.72328$ & $1.72328$ & $0.72214$ & $1.72214$\\
$0.89478$ & $1.89478$ & $1.10829$ & $2.10829$ & $0.97526$ & $1.97526$ & $0.89694$ & $1.89694$ & $0.89492$ & $1.89492$\\
$1.11685$ & $2.11685$ & $1.49495$ & $2.49495$ & $1.24518$ & $2.24518$ & $1.12022$ & $2.12022$ & $1.11706$ & $2.11706$\\
$1.38826$ & $2.38826$ & $2.02510$ & $3.02510$ & $1.57903$ & $2.57903$ & $1.39311$ & $2.39311$ & $1.38857$ & $2.38857$\\
$1.70903$ & $2.70903$ & $2.73639$ & $3.73639$ & $1.98079$ & $2.98079$ & $1.71562$ & $2.71562$ & $1.70944$ & $2.70944$\\
$2.07914$ & $3.07914$ & $3.66170$ & $4.66170$ & $2.24518$ & $3.24518$ & $2.08775$ & $3.08775$ & $2.07967$ & $3.07967$\\
$2.49859$ & $3.49859$ & $4.81699$ & $5.81699$ & $2.45636$ & $3.45636$ & $2.50950$ & $3.50950$ & $2.49927$ & $3.49927$\\
$2.96740$ & $3.96740$ & $6.18994$ & $7.18994$ & $3.01379$ & $4.01379$ & $2.98087$ & $3.98087$ & $2.96824$ & $3.96824$\\
$3.48556$ & $4.48556$ & $7.73254$ & $8.73254$ & $3.66340$ & $4.66340$ & $3.50185$ & $4.50185$ & $3.48657$ & $4.48657$\\
$4.05306$ & $5.05306$ & $9.35997$ & $10.3600$ & $4.41763$ & $5.41763$ & $4.07245$ & $5.07245$ & $4.05426$ & $5.05426$\\
$4.66991$ & $5.66991$ & $10.9570$ & $11.9570$ & $5.29075$ & $6.29075$ & $4.69267$ & $5.69267$ & $4.67132$ & $5.67132$\\
$5.33611$ & $6.33611$ & $12.3914$ & $13.3914$ & $6.29827$ & $7.29827$ & $5.36250$ & $6.36250$ & $5.33775$ & $6.33775$\\
$6.05165$ & $7.05165$ & $13.5314$ & $14.5314$ & $7.45618$ & $8.45618$ & $6.08195$ & $7.08195$ & $6.05354$ & $7.05354$\\
$6.81655$ & $7.81655$ & $14.2656$ & $15.2656$ & $8.78004$ & $9.78004$ & $6.85102$ & $7.85102$ & $6.81869$ & $7.81869$\\
\hline
\specialrule{.15em}{.15em}{.0em}
\end{tabular}
\end{center}
\end{table}

The number of degrees of freedom scales exponentially with resolution $k$ and linearly with the length $L$ of the interval. As the resolution truncation of the model is increased from $k=0$ to $k=1$ to $k=6$ to $k=10$, the number of degrees of freedom increases from $32$ to $72$ to $2552$ to $40952$. We see that the relative error in the lowest $\omega_p^2$, obtained through the numerical diagonalization of the coupling matrix $\Omega^2$, reduces from $2.19797\% (k=0)$ to $0.939075\% (k=1)$ to $0.0256624\% (k=6)$ to $ 0.00159811\% (k=10)$.

In what follows, we apply the flow equations to this model truncated to resolution $k=1$. To derive the flow equation, Eq. (\ref{eq:flow-equation_of_hamiltonian}), we apply the unitary transformation, $U(\lambda)$, parametrized by the continuous flow parameter $\lambda$, to the Hamiltonian described in Eq. (\ref{eq:psiphi_hamiltonian_matrix_form}). By following the procedure similar to one outlined in Sec. \ref{sec:The_scalar_field_theory_and_the_normal_mode_frequencies}, we derive the the flow equation) for the Hamiltonian,  
\begin{eqnarray}
\frac{d\mathrm{H}(\lambda)}{d\lambda}=\sum_{m,n=-\infty}^{\infty}\frac{1}{2}
\left[
\begin{pmatrix}
\phi^{s,k}_{m} & \psi^{s,k}_{m} & \phi^{w,l}_{m} & \psi^{w,l}_{m}
\end{pmatrix}
\left[K(\lambda),\Omega^2(\lambda)\right]
\begin{pmatrix}
\phi^{s,k}_{n} \\ \psi^{s,k}_{n} \\ \phi^{w,q}_{n} \\ \psi^{w,q}_{n}
\end{pmatrix}
\right].
\end{eqnarray}
Here, $\Omega^2(\lambda)=O^T(\lambda) \Omega^2 O(\lambda)$, is the equivalent matrix of $\Omega^2$ at an arbitrary value of the flow parameter $\lambda$. $K(\lambda)$ is the generator of transformation, which is chosen to be of the same type as that presented in Eq. (\ref{eq:the_generator_3}). So, the flow equation of the full Hamiltonian reduces to the form given in Eq. (\ref{eq:the_flow_equation_omega}).

\begin{figure}[hbt]
\begin{center}
\caption{The schematic diagram of the scale coupling matrix, $\Omega^2$, for the interaction of two scalar fields in the theory with truncated resolution 1.}
\label{fig:the_structural_composition_of_omega_square}
\includegraphics[scale=0.40]{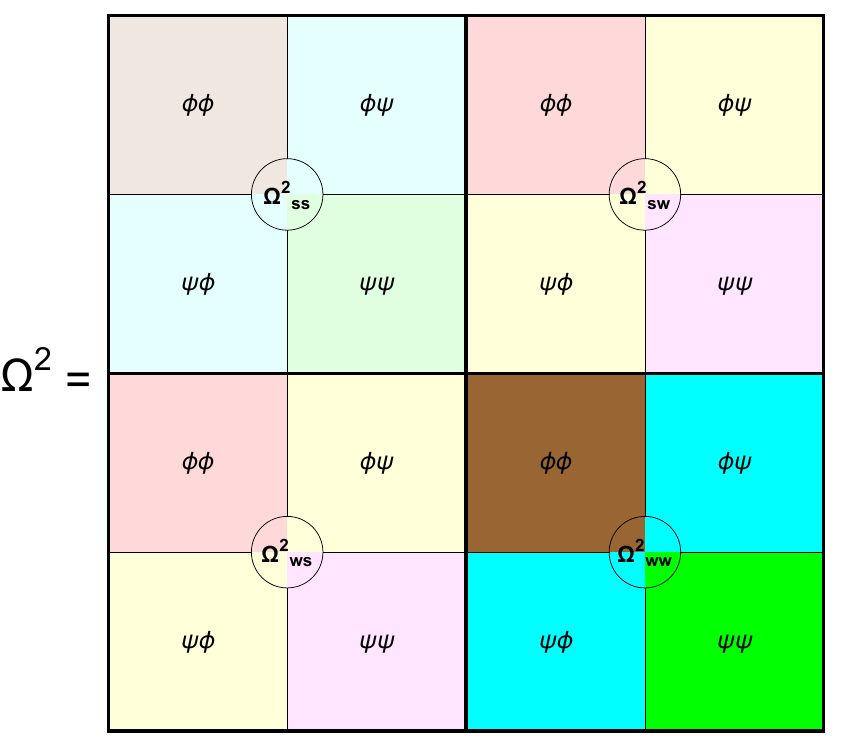}
\end{center}
\end{figure}

The flow equations are implemented for mass parameters $\mu$, $\nu$, and $g$ set to $1$. The Hamiltonian is truncated to subspace $\mathcal{H}^1$ visualized as a direct sum of resolution $0$ scaling subspace $\mathcal{H}^0$ and resolution $0$ wavelet subspace $\mathcal{W}^0$. Furthermore, we limit the translation index such that the basis elements are fully contained within $x=0$ and $x=20$. Such a truncation requires $16$ scaling and $16$ wavelet function modes for analysis. The resulting form of the coupling matrix $\Omega^2$ truncated to resolution $k=1$ is visualized, as shown in Fig. \ref{fig:the_structural_composition_of_omega_square}.
 
The truncated matrix is composed of four major blocks labelled as $\Omega^2_{ss}$, $\Omega^2_{sw}$, $\Omega^2_{ws}$, and $\Omega^2_{ww}$. The subscripts $ss$ and $ww$ indicate that these blocks represent couplings within the $k=0$ scaling and $k=0$ wavelet spaces, respectively. In contrast, the blocks labelled by subscripts $sw$ and $ws$ signify that they consist of coupling coefficients between the $k=0$ scaling and $k=0$ wavelet spaces. Each of these major matrix blocks contains four minor blocks, labelled as $\phi\phi$, $\phi\psi$, $\psi\phi$, and $\psi\psi$. The sub-blocks $\phi\phi$ and $\psi\psi$ contain couplings associated with the $\phi$ or $\psi$ fields, respectively, while $\phi\psi$ and $\psi\phi$ contain coupling terms between the two. The normal mode frequencies computed by numerical diagonalization of the truncated $\Omega^2$ are presented in the first two columns of Table. \ref{tab:normal_mode_frequencies}. It is worth noting that these values differ slightly from those presented in the $k=1$ column of Table. \ref{tab:compare_omega_and_omega_for_resolution_10}. This is because the nature of truncation slightly differs when applied to the coupling matrix constructed in scaling space $\mathcal{H}^1$ as against the one constructed in $\mathcal{H}^0\oplus \mathcal{W}^0$.

The results emerging from the application of the flow equation to the coupling matrix $\Omega^2$ are presented in Table. \ref{tab:normal_mode_frequencies} at various stages of the flow. For each value of $\lambda$, the normal mode frequencies associated with the renormalized fields $\Phi$ and $\Psi$ are computed by diagonalizing the $\Omega^2_{ss}$. Inspection of each row of Table. \ref{tab:normal_mode_frequencies} shows that the low-lying eigenvalues approach the exact eigenvalues of the truncated theory. 
\begin{table}[hbt]
\begin{center}
\caption{Comparison among the $\omega_p^2$ governed by the exact diagonalization the low-lying eigenvalues approaches towards the exact eigenvalues of the truncated theory. of the truncated $\Omega^2$ matrix and the $\Omega^2_{ss}$ part of the matrix for different values of the flow parameter, $\lambda$}
\label{tab:normal_mode_frequencies}
\setlength{\tabcolsep}{0.45pc}
\vspace{1mm}
\begin{tabular}{c | c | c | c | c | c | c | c | c | c | c | c}
\specialrule{.15em}{.0em}{.15em}
\hline
\multicolumn{2}{c|}{Truncated exact} & \multicolumn{2}{c|}{$\lambda$=0} & \multicolumn{2}{c|}{$\lambda$=0.1} & \multicolumn{2}{c|}{$\lambda$=1} & \multicolumn{2}{c|}{$\lambda$=5} & \multicolumn{2}{c}{$\lambda$=20}\\
\hline
$\Phi$ & $\Psi$ & $\Phi$ & $\Psi$ & $\Phi$ & $\Psi$ & $\Phi$ & $\Psi$ & $\Phi$ & $\Psi$ & $\Phi$ & $\Psi$\\
\hline
$0.53595$ & $1.53595$ & $0.53621$  & $1.53621$ & $0.53595$ & $1.53595$ & $0.53595$ & $1.53595$ & $0.53595$ & $1.53595$ & $0.53595$ & $1.53595$\\
$0.64408$ & $1.64408$ & $0.64567$  & $1.64567$ & $0.64408$ & $1.64408$ & $0.64408$ & $1.64408$ & $0.64408$ & $1.64408$ & $0.64408$ & $1.64408$\\
$0.82537$ & $1.82537$ & $0.83255$  & $1.83255$ & $0.82537$ & $1.82537$ & $0.82537$ & $1.82537$ & $0.82537$ & $1.82537$ & $0.82537$ & $1.82537$\\
$1.08189$ & $2.08189$ & $1.10829$  & $2.10829$ & $1.08189$ & $2.08189$ & $1.08189$ & $2.08189$ & $1.08189$ & $2.08189$ & $1.08189$ & $2.08189$\\
$1.41694$ & $2.41694$ & $1.49495$  & $2.49495$ & $1.41694$ & $2.41694$ & $1.41694$ & $2.41694$ & $1.41694$ & $2.41694$ & $1.41694$ & $2.41694$\\
$1.83517$ & $2.83517$ & $2.02510$  & $3.02510$ & $1.83517$ & $2.83517$ & $1.83517$ & $2.83517$ & $1.83517$ & $2.83517$ & $1.83517$ & $2.83517$\\
$2.34243$ & $3.34243$ & $2.73639$  & $3.73639$ & $2.34243$ & $3.34243$ & $2.34243$ & $3.34243$ & $2.34243$ & $3.34243$ & $2.34243$ & $3.34243$\\
$2.94602$ & $3.94602$ & $3.66170$  & $4.66170$ & $2.94602$ & $3.94602$ & $2.94602$ & $3.94602$ & $2.94602$ & $3.94602$ & $2.94602$ & $3.94602$\\
$3.65564$ & $4.65564$ & $4.81698$  & $5.81698$ & $3.65564$ & $4.65564$ & $3.65564$ & $4.65564$ & $3.65564$ & $4.65564$ & $3.65564$ & $4.65564$\\
$4.48633$ & $5.48633$ & $6.18994$  & $7.18994$ & $4.48633$ & $5.48633$ & $4.48633$ & $5.48633$ & $4.48633$ & $5.48633$ & $4.48633$ & $5.48633$\\
$5.46196$ & $6.46196$ & $7.73254$  & $8.73254$ & $5.46196$ & $6.46196$ & $5.46196$ & $6.46196$ & $5.46196$ & $6.46196$ & $5.46196$ & $6.46196$\\
$6.61455$ & $7.61455$ & $9.35997$  & $10.3560$ & $6.61455$ & $7.61455$ & $6.61455$ & $7.61455$ & $6.61455$ & $7.61455$ & $6.61455$ & $7.61455$\\
$7.97792$ & $8.97792$ & $10.9570$ & $11.9570$ & $7.97792$ &  $8.97792$ & $7.97792$ & $8.97792$ & $7.97792$ & $8.97792$ & $7.97792$ & $8.97792$\\
$9.58140$ & $10.5814$ & $12.3914$ & $13.3914$ & $9.58140$ &  $10.5814$ & $9.58140$ & $10.5814$ & $9.58140$ & $10.5814$ & $9.58140$ & $10.5814$\\
$11.4468$ & $12.4468$ & $13.5314$ & $14.5314$ & $11.4796$ &  $12.4796$ & $11.4468$ & $12.4468$ & $11.4468$ & $12.4468$ & $11.4468$ & $12.4468$\\
$13.5869$ & $14.5869$ & $14.5314$ & $15.5314$ & $13.6997$ &  $14.6997$ & $13.5869$ & $14.5869$ & $13.5869$ & $14.5869$ & $13.5869$ & $14.5869$\\
\hline
\specialrule{.15em}{.15em}{.0em}
\end{tabular}
\end{center}
\end{table}
For example, the order of the difference between the flow equation computed and the truncated exact lowest eigenvalue reduces from $10^{-3}$ to $10^{-7}$ to $10^{-12}$ as $\lambda$ flows from $0$ to $1$ to $20$. This means that the flow equations are able to incorporate the effects of finer resolution into the coarsest resolution sector. 

\begin{figure}[H]
\begin{center}
\caption{Hilbert-Schmidt norm of all four types of non-zero quadratic expressions as a function of the flow parameter.}
\label{fig:HS_norm_of_four_types_of_quadratic expressions}
\includegraphics[scale=0.40]{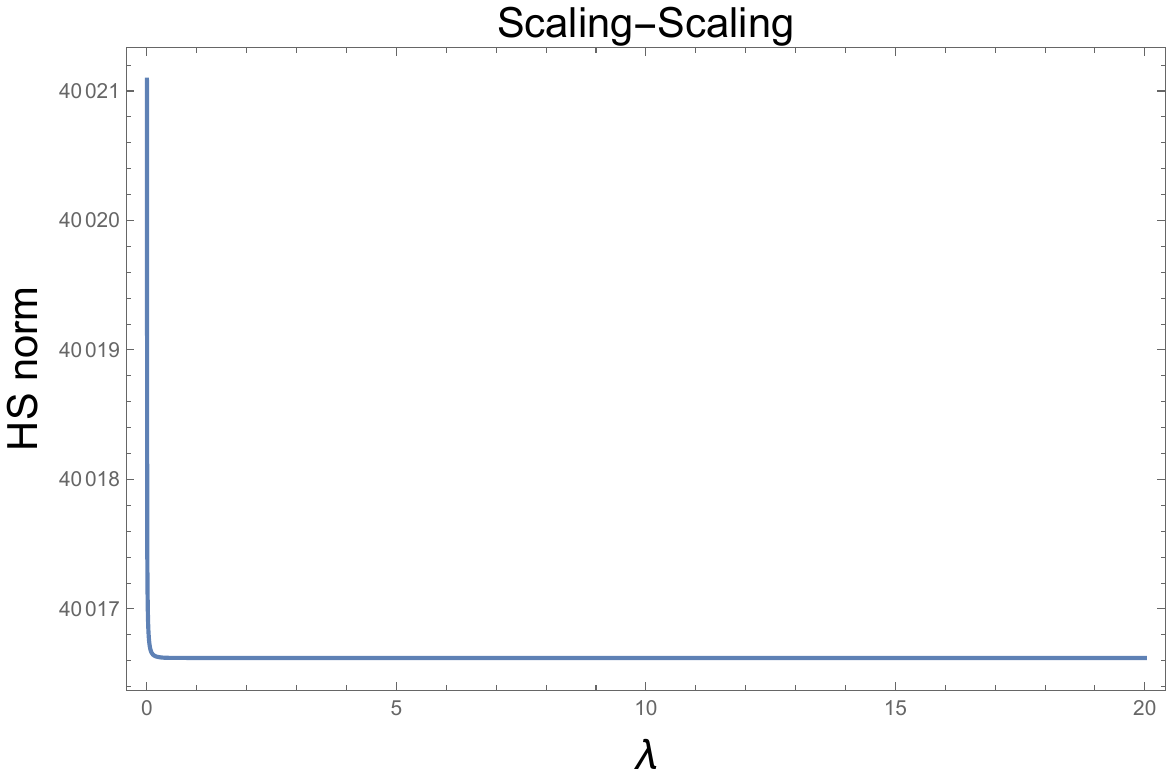}
\includegraphics[scale=0.388]{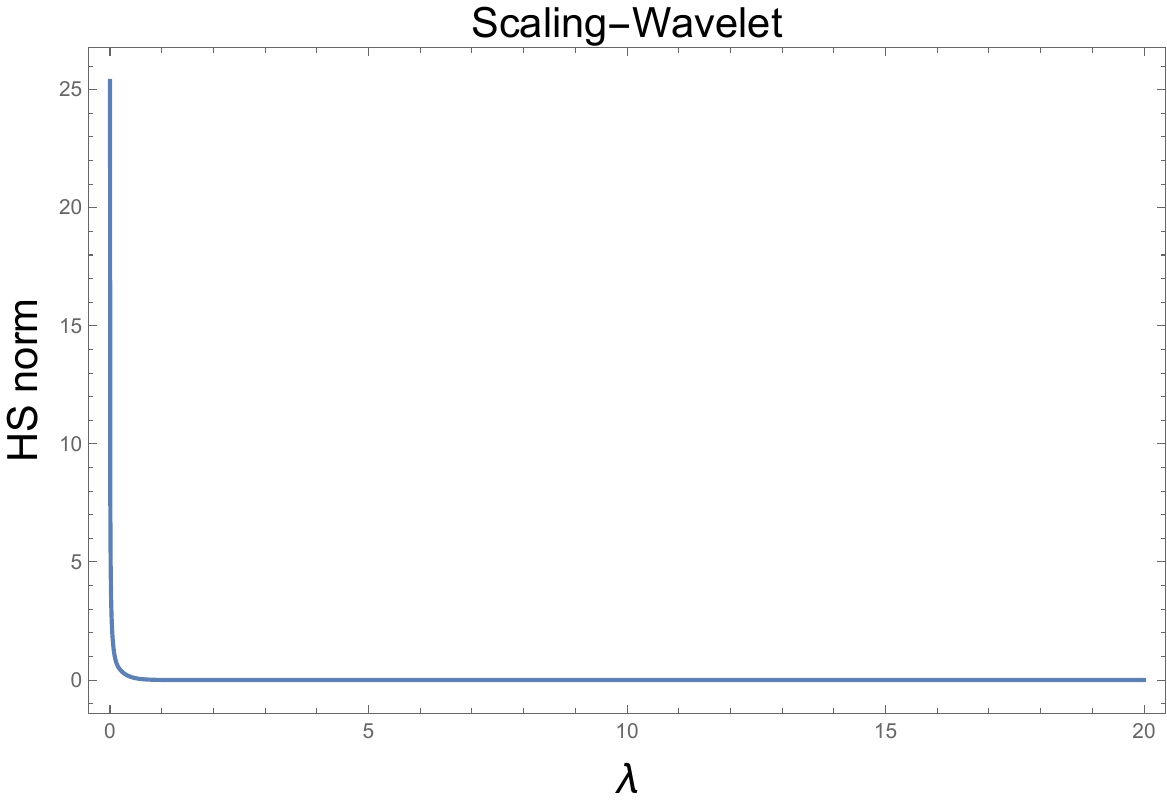}
\includegraphics[scale=0.388]{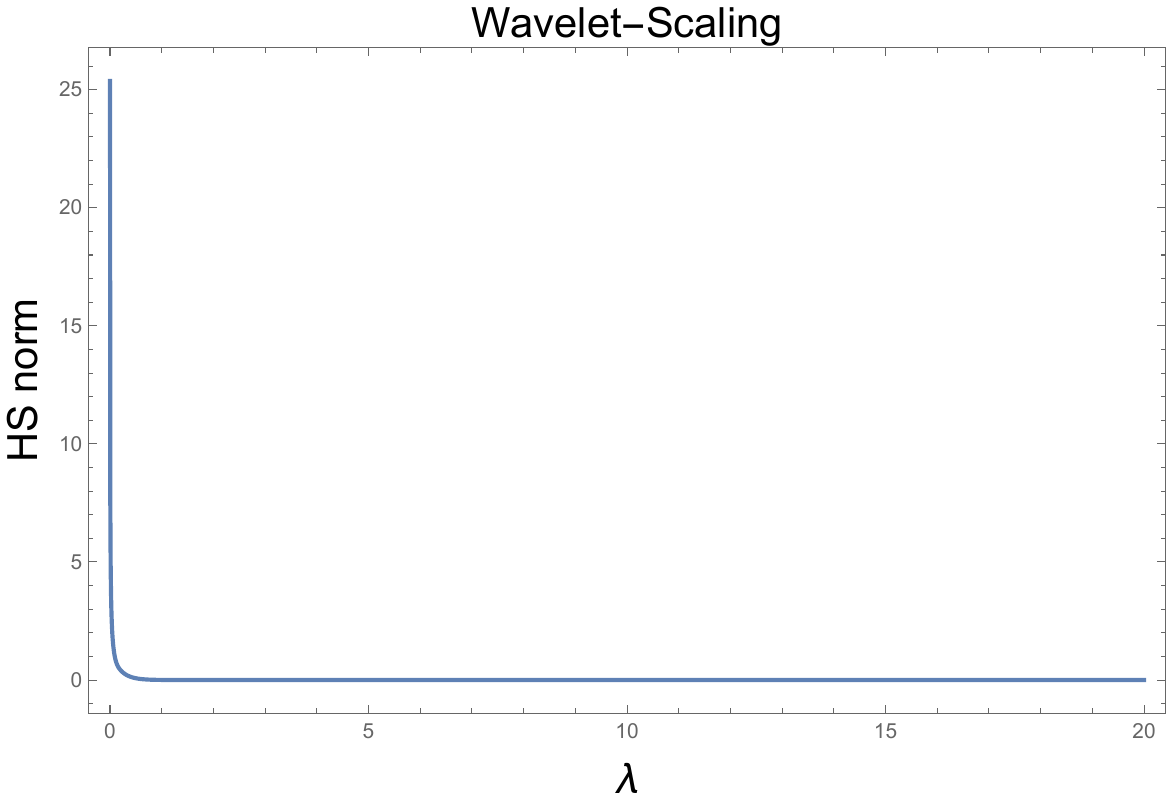}
\includegraphics[scale=0.40]{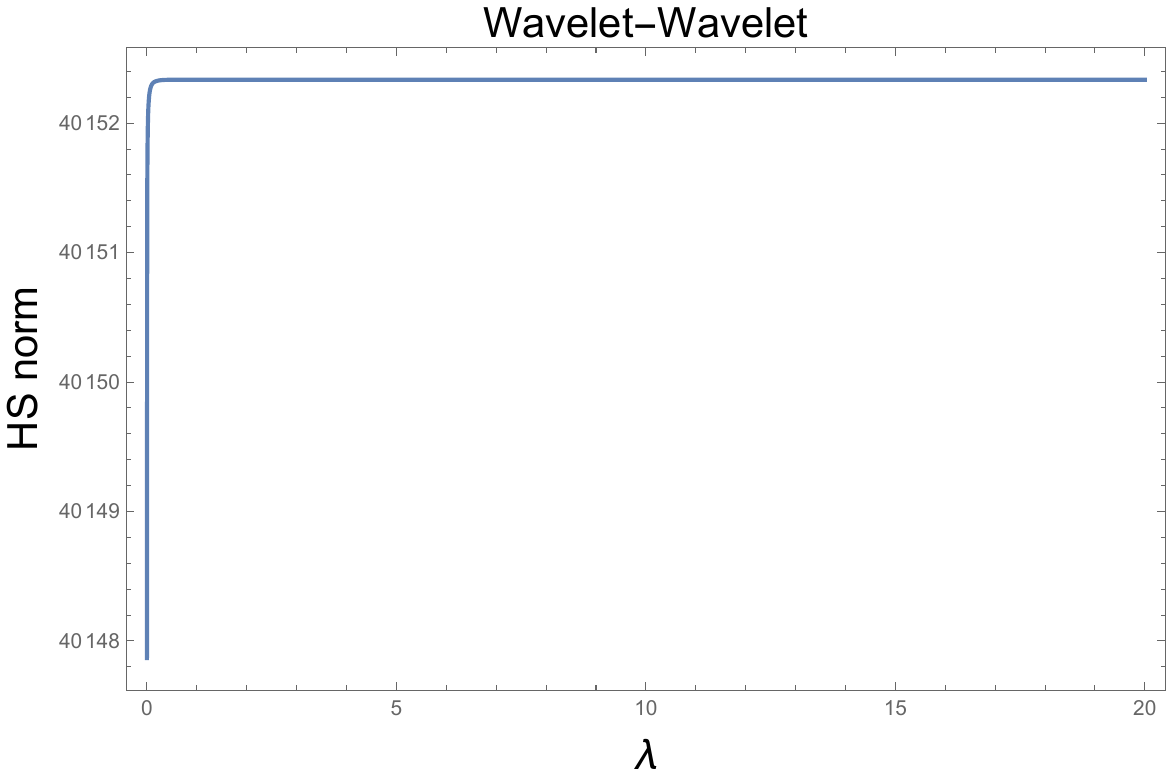}
\end{center}
\end{figure}

The decoupling of degrees of degrees of freedom associated with different scales or resolutions is validated through the computation of the HS norms. These HS norms, defined by Eq. (\ref{eq:Hilbert-Schmidt_norm_1}), for each of the major blocks, $\Omega^{2}_{ss}$, $\Omega^{2}_{sw}$, $\Omega^{2}_{ws}$, and $\Omega^2_{ww}$, are plotted in Fig. \ref{fig:HS_norm_of_four_types_of_quadratic expressions} as a function of the flow-parameter, $\lambda$. 
We observe that the HS norms for the intra-scale coupling blocks saturate to nontrivial values, whereas HS norms for the inter-scale coupling blocks approach zero as the value of $\lambda$ increases. This behaviour of the flow of the HS norm is analogous to the nature of the HS norm obtained for scalar field theory. This resemblance suggests that the flow equation method can be useful to derive an equivalent Hamiltonian, consisting of physically relevant degrees of freedom within the context of the theory of two interacting scalar fields.

As the HS norm is dominated by the largest matrix elements, it is also useful to understand how the individual matrix elements will evolve with the increasing value of $\lambda$. To illustrate this, we presented the plot of the individual matrices for the different values of $\lambda$ in Fig. \ref{fig:HS_norm_of_four_types_of_quadratic expressions}. 
\begin{figure}[hbt]
\begin{center}
\label{fig:matrix_plot_for_different_values_of_lambda}
\caption{The matrix plots of the scale coupling matrix, $\Omega^2$, with the increasing value of $\lambda$.}
\includegraphics[scale=0.35]{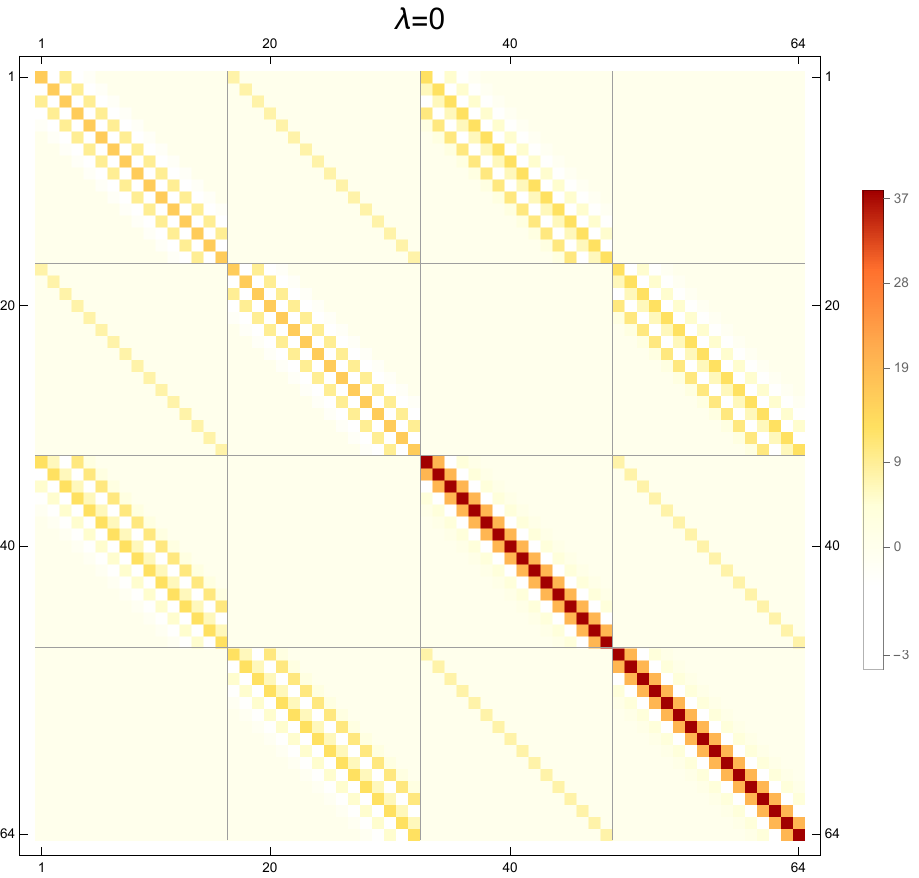}
\includegraphics[scale=0.35]{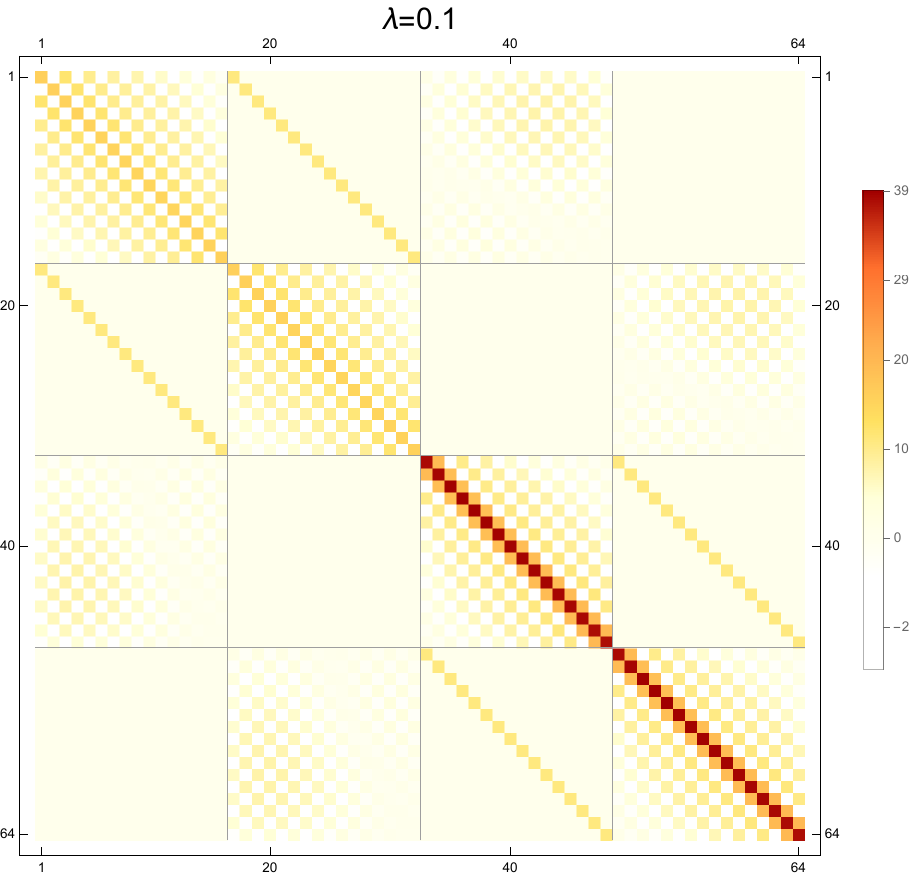}
\includegraphics[scale=0.35]{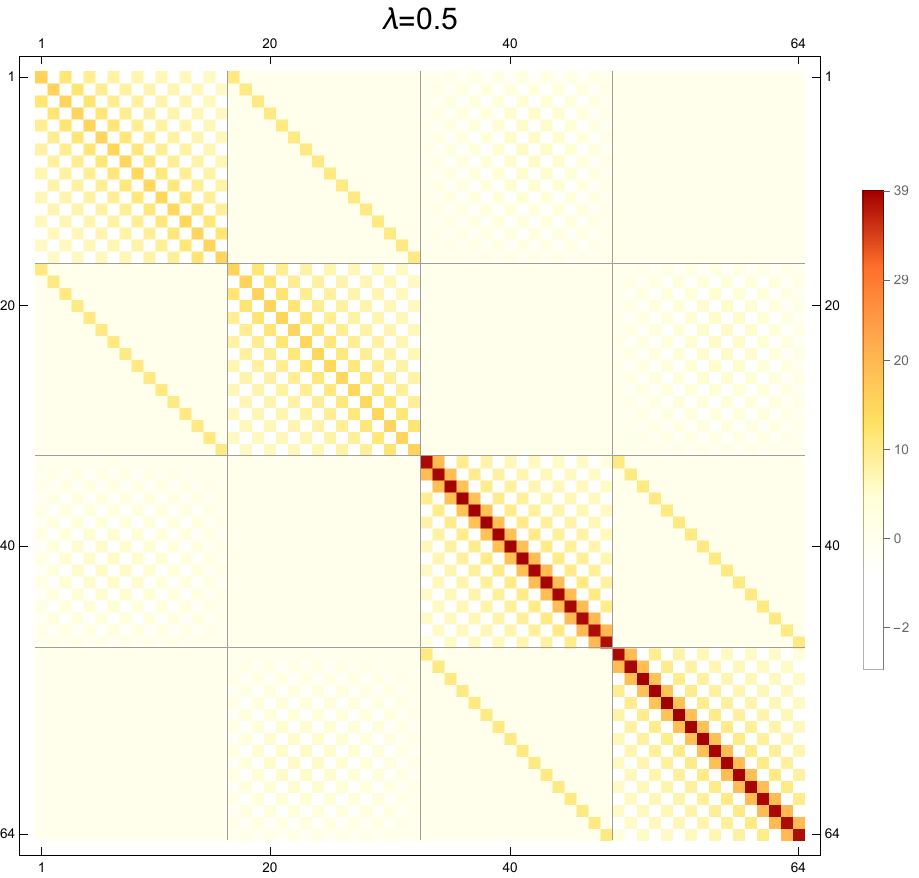}
\includegraphics[scale=0.35]{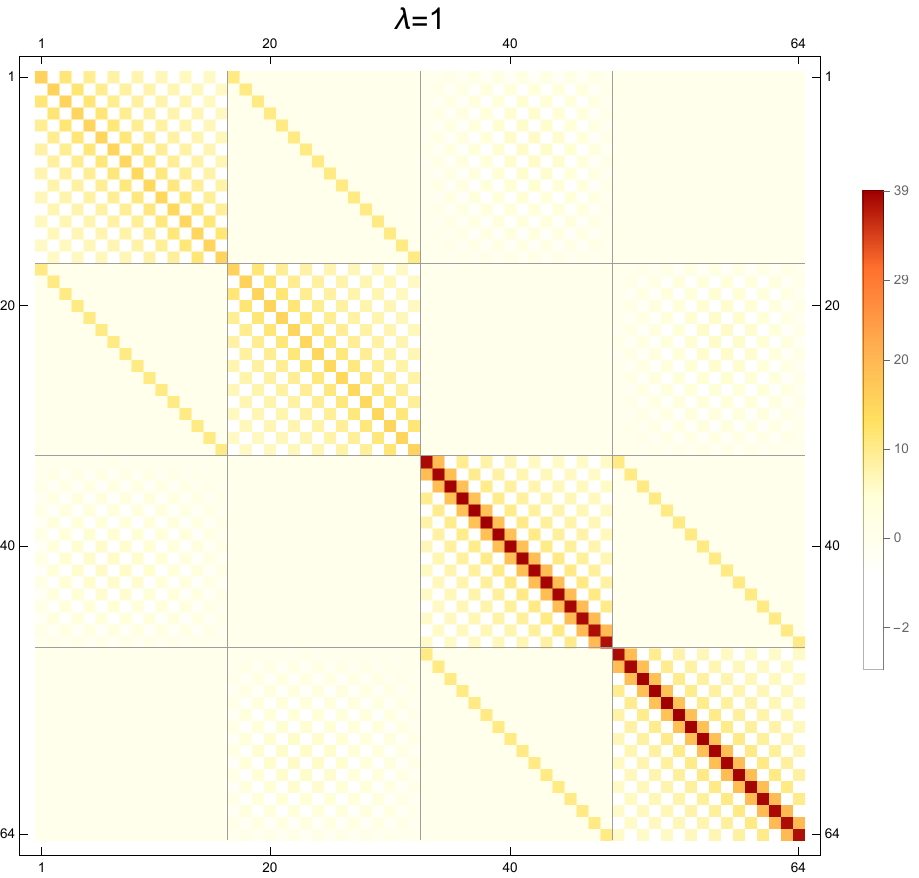}
\includegraphics[scale=0.35]{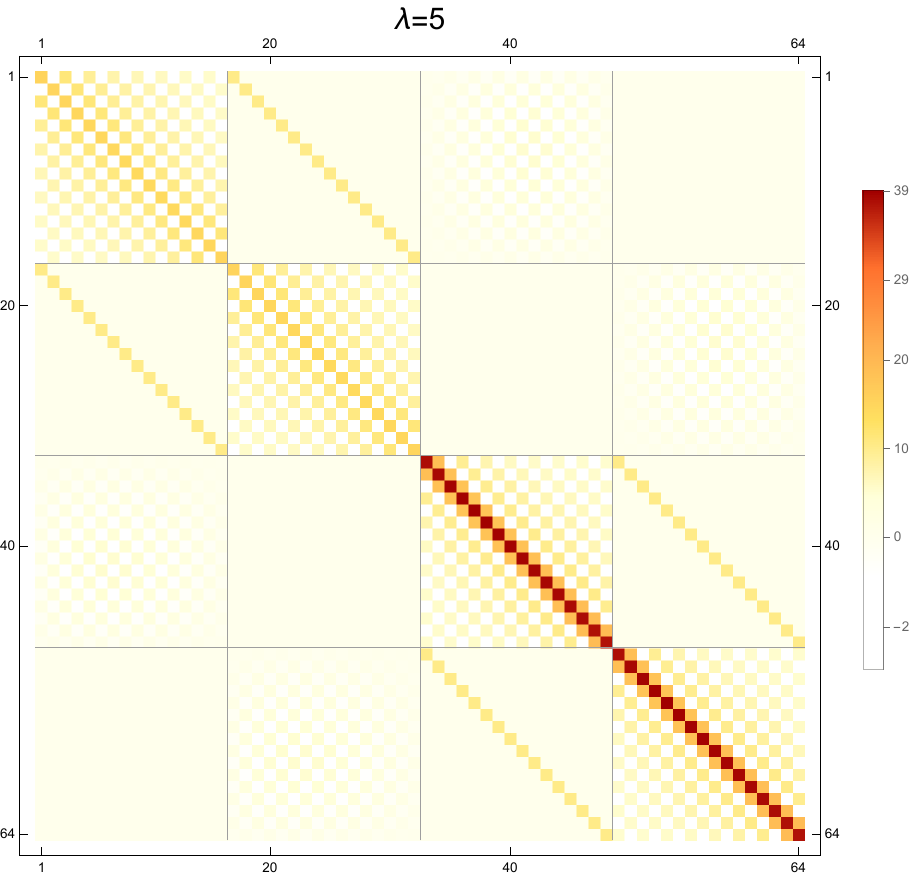}
\includegraphics[scale=0.35]{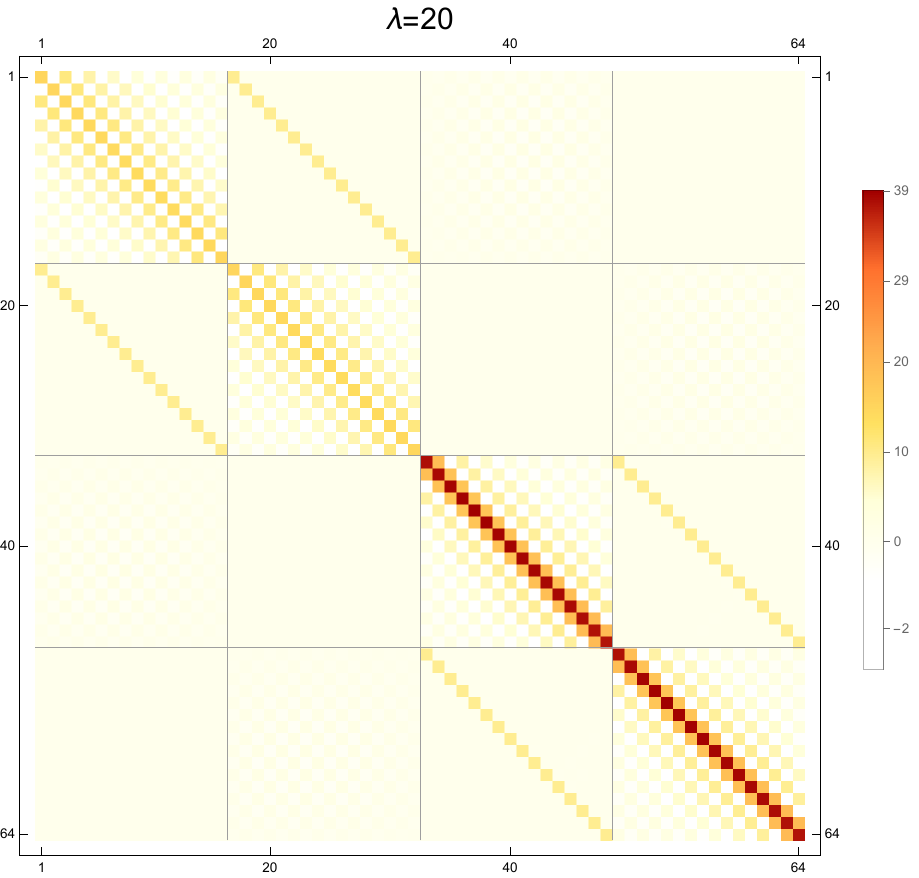}
\end{center}
\end{figure}
The constitution of each of the figures is similar to the schematic diagram, depicted in Fig. \ref{fig:the_structural_composition_of_omega_square}. There are $16$ small matrix blocks in each matrix plots. Each matrix block is $16 \times 16$ matrix.

The first plot, in Fig. \ref{fig:matrix_plot_for_different_values_of_lambda}, represents the matrix before the flow is commenced, i.e. for $\lambda=0$. Here, The diagonal blocks, $\Omega^2_{ss}$ and $\Omega^2_{ww}$, are band diagonal as only a few numbers of overlap integrals, $\mathcal{D}^k_{ss,mn}$ and $\mathcal{D}^k_{ww,mn}$, exists. Similarly, due to the existence of a limited number of overlap integrals, $\mathcal{D}^{kq}_{sw,mn}$ and $\mathcal{D}^{lk}_{ws,mn}$, the off-diagonal matrix blocks, $\Omega^2_{sw}$ and $\Omega^2_{ws}$, are also band diagonal. It can be observed in the figure that the off-diagonal blocks, labelled as $\phi\psi$ and $\psi\psi$, inside the bigger diagonal blocks ($\Omega^2_{ss}$ and $\Omega^2_{ww}$), only the diagonal elements are non-zero. These diagonal elements represent the coupling between the two fields $\phi$ and $\psi$ on the same scale. It's noticeable that the different scale coupling between two fields, $\phi$ and $\psi$, contained in the off-diagonal blocks of $\Omega^2_{sw}$ and $\Omega^2_{ws}$, $\phi\psi$ and $\psi\phi$, are equal to zero.

It can be seen from the subsequent figures that as the value of $\lambda$ begins to increase from zero, the coupling terms between different scales start to decrease and ultimately become zero. For $\lambda=0.1$ and $\lambda=0.5$, the different scale coupling terms are weaker as compared to the coupling terms corresponding to $\lambda=0$, but not zero. At $\lambda=1$, the off-diagonal terms become almost zero. The value of the off-diagonal elements does not change significantly with the increasing values of $\lambda$ beyond $\lambda=1$. At $\lambda=20$, the matrix is almost block diagonal. More specifically, as seen in the plot, each diagonal matrix remains band diagonal, even after the flow commences, but the band gets widened. This phenomenon signifies the creation of new coupling terms in the effective Hamiltonian, which includes the effect of the eliminated short-distance degrees of freedom. The band diagonal structure of the new effective matrix, $\Omega^2_{ss}$, signifies the preservation of the local nature of the theory. 

In the case of two interacting scalar fields, the flow equation can decouple the coarser and the finer scale degrees of freedom. At each value of the flow parameter, the flow equation will produce a new equivalent Hamiltonian. Ignoring the coupling terms at each value of $\lambda$, the Hamiltonian is a sum of two operators associated with coarser and finer scale degrees of freedom. But those degrees of freedom will include the effects of the eliminated coupled degrees of freedom.

\section{Conclusion and outlook}
\label{sec:conclusion}
In this work, we first studied scalar field theories using discrete wavelet-based flow equations. The noninteracting scalar field theory in $1+1$ dimensions is analysed at one resolution higher than previously studied \cite{PhysRevD.95.094501}. Within the discrete wavelet framework, the Hamiltonian has a natural multi-block structure with diagonal blocks representing couplings across locations at a given resolution, while the off-diagonal blocks represent couplings between different resolutions. The application of flow equations at resolution $k=2$ reveals that the Hamiltonian flows to a block diagonal form, thereby decoupling interactions between resolutions. This approach will be beneficial in situations wherein multiple resolutions compete and contribute to observables and correlation functions.

The second study analysed the model of two scalar fields interacting through a generally quadratic interaction. The application of the flow equations shows that the flow equations filter the interactions between the two fields at each resolution. At the same time, the interactions across resolutions between the same fields, as well as those between the two fields, get decoupled with the flow. The decoupling between resolutions shows up as an effective contribution to matrix elements within each diagonal block.   

In the above studies, we assumed that the flow neither induces changes in the terms involving canonical momenta nor does it induce terms that couple the canonical momenta with canonical coordinates. The results reported in \cite{PhysRevD.95.094501} justify this assumption and substantially reduce the computational effort.  

The quantum theory of a free field, as well as that of free fields coupled quadratically, offer exact analytical solutions. They provide a test bed for estimating errors in computations based on truncated theories in comparison to the exact analytical solutions. There are two kinds of truncation that arise when solving the Hamiltonian eigenvalue problem in quantum field theory. The first type of truncation sets a bound on the number of degrees of freedom in the Hamiltonian. In the wavelet-based QFT, this implies a volume and resolution truncation. The second type is Hilbert space truncation, which involves truncating the basis of the Hilbert space by, say, limiting the maximum energy of any basis state. The Hamiltonian matrix is rendered to a finite-dimensional form amenable to further treatment via numerical diagonalization. 

The truncation of field degrees of freedom in wavelet-based theories by volume and resolution, when applied to the operators representing the generators of Poincaré symmetry, will lead to the breaking of Poincaré invariance. The wavelet-based flow equation approach that renders the Hamiltonian block diagonal by resolution will likely flow these generators to more complicated forms so that the effects of the breaking of Poincaré invariance reduces with the flow. The manner in which this occurs in worth investigating.

Another problem that deserves attention concerns the dynamics of the coupled evolution of two-level systems interacting with a real or complex scalar field from a wavelet perspective. This problem, as indicated in the introduction, was the one investigated by Wilson for which the idea of wavelets was first hypothesized.  

The use of light front in the study of QCD has been advocated as an important element in reconciling the current quark picture with the constituent picture that emerges in the low energy sector. The non-covariant nature of light front formulation leads to a complex structure of ultraviolet and infrared divergences. The structure of these divergences was estimated using a phase space cell analysis based on the concept of wavelets. However, the analysis was qualitative in nature and merits reworking within the framework of Daubechies wavelets.

\appendix

\bibliographystyle{unsrt}
\bibliography{References}
\end{document}